\documentclass[11pt]{article}
\usepackage[utf8]{inputenc}
\usepackage[a4paper, margin=2.5cm]{geometry}
\usepackage{graphicx}
\usepackage{xcolor}
 \usepackage{float} 
\usepackage{latexsym}
\usepackage{comment}
\usepackage{multicol}
\usepackage{dsfont}
\usepackage{braket}
\usepackage{amssymb,amsthm, amsmath}
\usepackage{quantikz}
\usepackage{hyperref}
\usepackage{subcaption}
\usepackage{caption}
\usepackage{mathtools} 
\usepackage{dsfont}
\usepackage{booktabs}
\usepackage{array}
\usepackage[backend=bibtex,sorting=none]{biblatex}
\usepackage{soul}
\usepackage{authblk}

\def\ba{{\bf a}}
\def\bb{{\bf b}}
\newcommand{\bR}{{\mathbb R }}

\newcommand{\al}{{|\alpha|}}
\newcommand{\be}{{|\beta|}}
\newcommand{\ga}{{|\gamma|}}
\newcommand{\de}{{|\delta|}}
\newcommand{\Tr}{{\mathrm Tr}}
\newcommand{\btheta}{\boldsymbol{\theta}}
\newcommand{\bphi}{\boldsymbol{\phi}}

\newtheorem{theorem}{Theorem}

\definecolor{yellow_sdrac}{RGB}{255,238,175}

\title{A resource-efficient and noise-robust entanglement witness based on the swap test}

\author[1]{ Sebastiano Guaraldo }
\author[2,3]{ Sonia Mazzucchi }
\author[1]{ Alessio Baldazzi }
\author[1]{ Stefano Azzini }
\author[1]{ Lorenzo Pavesi }

\affil[1]{Department of Physics, University of Trento, Trento, 38123, Italy}
\affil[2]{Department of Mathematics, University of Trento, Trento, 38123, Italy}
\affil[3]{Istituto Nazionale di Fisica Nucleare -  Trento Institute for Fundamental Physics and Applications (INFN-TIFPA), Trento, 38123, Italy}

\date{}

\begin{document}

\maketitle

\abstract{Quantum entanglement is an essential resource for quantum technologies, and the controlled swap test provides a versatile tool for its detection and quantification. Here, we propose a SWAP-based entanglement witness that applies to arbitrary two-qubit states - both pure and mixed - and provides a lower bound on the concurrence. The method is resource-efficient, robust to noise, and platform-independent. As an example, we validate the approach on a room-temperature photonic chip, where the swap test is carried out using only linear and well-established integrated optical components. The robustness of the method against photonic-hardware noise is also analysed. Our results establish a simple and reliable tool for entanglement witnessing.}

\section{Introduction}
\label{sec:intro}
Entanglement is a fundamental and unique feature of quantum mechanics \cite{einstein1935can} that cannot be replicated by classical systems, as demonstrated by Bell's theorem \cite{bell1964einstein, brunner2014bell}. Nowadays, it is recognised as a fundamental resource \cite{horodecki2009quantum} in quantum computing \cite{nielsen2010quantum, jozsa2003role, ekert1998quantum}, quantum communication \cite{bennett1992communication, bennett1993teleporting, ekert1991quantum} and quantum information \cite{keyl2002fundamentals}, as well as in quantum sensing \cite{degen2017quantum, giovannetti2011advances} and quantum technologies more in general \cite{zhang2024entanglement}. However, identifying and quantifying entanglement remains a challenging task. In fact, determining whether a given quantum state is separable or entangled - known as the \textit{separability problem} - has been shown to be NP-hard \cite{gurvits2003classical}. Nevertheless, several criteria have been developed to detect entanglement, such as the positive partial transpose (PPT) criterion \cite{peres1996separability, Horodecki:1996nc} and the computable cross-norm or realignment (CCNR) criterion \cite{chen2002matrix, rudolph2004computable}. These methods require full knowledge of the system’s density matrix, which must be reconstructed through quantum state tomography \cite{horodecki2009quantum}. Since the number of measurements needed scales exponentially with the Hilbert space dimension, tomography quickly becomes infeasible as the system size increases \cite{altepeter20044, gross2010quantum, cramer2010efficient}. While advanced approaches such as compressed-sensing tomography \cite{PhysRevLett.105.150401} and threshold quantum state tomography \cite{caruccio2025experimental, PhysRevA.111.032436} can significantly reduce the number of required measurements, they still rely on partial state reconstruction and may impose additional assumptions on the state under investigation. Moreover, in realistic experimental settings, additional complications arise. Access to the full set of measurement bases required for tomography may be experimentally demanding or limited by technical constraints \cite{friis2019entanglement, imai2024collective, haffner2005scalable}. Environmental noise and decoherence further degrade quantum correlations, reducing the purity and detectability of entanglement \cite{zurek2003decoherence, schlosshauer2019quantum}. This inherent fragility makes entanglement highly sensitive to experimental imperfections, making its detection particularly challenging in noisy intermediate-scale quantum (NISQ) devices \cite{preskill2018quantum}. In this context, entanglement witnesses offer a powerful alternative \cite{guhne2009entanglement}. They are designed to distinguish certain entangled states from all separable ones, without the need for full state reconstruction. Although they cannot detect all entangled states, witnesses are highly useful in experimental and applied contexts, where efficiency and scalability are crucial. Moreover, in many practical scenarios - such as quantum teleportation \cite{bennett1993teleporting} or quantum key distribution \cite{ekert1991quantum} - only a specific class of entangled states is of interest. In these cases, an observable can be tailored to identify that class precisely, without the need of universal detection and with the possibility of improving simplicity and efficiency. 

Foulds et al. \cite{foulds2021controlled} proposed a method to detect and quantify entanglement in multipartite pure states using a parallelized swap test \cite{barenco1997stabilization, buhrman2001quantum}. Their approach requires two copies of the input state and one auxiliary control qubit per system qubit. For instance, in the two-qubit case, four Hadamard (H) gates and two controlled-SWAP\footnote{Throughout this work, ‘SWAP’ in capital letters refers to the quantum gate that implements the swap operation.} (CSWAP) gates are employed. In their protocol, the probability of measuring both control qubits in the state $\ket{1}$ is directly related to the concurrence, a standard bipartite entanglement measure ranging from $0$ for separable states to $1$ for maximally entangled ones \cite{wootters2001entanglement}. Subsequent theoretical works \cite{beckey2021computable, foulds2024generalizing, zhang2024controlled, wang2025efficient} have extended SWAP-based entanglement estimation to multipartite systems and mixed states. Specifically, Beckey \textit{et al.} \cite{beckey2021computable} introduced a family of multipartite entanglement measures, called \textit{concentratable entanglements}, which not only quantify global entanglement but also characterize the entanglement within and between subsystems of a composite quantum state. For pure states, these quantities can be estimated via the $n$-qubit parallelized swap test. Zhang \textit{et al.} \cite{zhang2024controlled} generalized the work of Foulds \textit{et al.} \cite{foulds2021controlled} to arbitrary mixed states. They also introduced a criterion for genuine multipartite entanglement in pure states and derived a lower bound on concentratable entanglement measures for multipartite mixed states, both inferred from the control-qubit statistics of the swap test circuit.

Although powerful, such protocols remain experimentally demanding, as they rely on multiple auxiliary qubits, several CSWAP gates, and duplicated copies of the input state. In this work, we show that, in contrast, the standard two-qubit swap test - with a single ancilla and a single CSWAP gate - already functions as an entanglement witness. By analysing the output statistics of the ancillary qubit, we also derive a lower bound on the concurrence of any two-qubit pure state. Moreover, since this bound is a convex function of the probability of measuring the ancilla in the state $\ket{1}$, the result can be extended to mixed states, making the protocol applicable to arbitrary two-qubit density matrices. This establishes the swap test as a resource-efficient witness: it extracts less information than full concurrence-estimation schemes but requires significantly fewer quantum resources. The protocol is platform-independent and can be implemented wherever a swap test can be carried out. We demonstrate it experimentally on a reconfigurable linear photonic integrated circuit (PIC) \cite{baldazzi2024linear}, validating the witness on both pure and mixed input states. We further analyse the robustness of the method under realistic noise, showing that the witness remains reliable even in the presence of phase fluctuations and imperfections in integrated optical components. Although our noise model is tailored to photonic hardware, the same methodology can be readily adapted to other architectures. Overall, these results establish the swap test as a practical and resource-efficient tool for detecting entanglement in two-qubit systems.

The paper is structured as follows. Sec.~\ref{sec:section_2} recalls the swap test algorithm and illustrates how it can be used to witness entanglement. Sec.~\ref{sec:section_3} describes the photonic implementation of the algorithm and analyzes the robustness of the approach against noise and circuit non-idealities. In Sec.~\ref{sec:section_4}, we present and discuss the experimental results, evaluating the circuit’s performance in identifying entanglement. Finally, Sec.~\ref{sec:section_5} summarizes the proposed method and key findings, discussing the advantages and limitations of the protocol and its photonic implementation.
\section{The swap test to witness entanglement}
\label{sec:section_2}
\begin{figure}
\centering
\begin{subfigure}[c]{0.49\textwidth}
\hspace*{-1.5cm}
\centering
\begin{quantikz}[row sep={0.6cm,between origins}, column sep=0.5cm]
\lstick{$\ket{0}$}      & \gate[style={fill=yellow_sdrac}]{\text{H}} & \ctrl{2}\gategroup[3,steps=1,style={dashed,rounded
 corners,fill=blue!15, inner
 xsep=4pt},background,label style={label
 position=below,anchor=north,yshift=-0.2cm}]{CSWAP}  & \gate[style={fill=yellow_sdrac}]{\text{H}} & \meter{} \\
\lstick{$\ket{\psi}$}   & \qw      & \targX{}   & \qw      & \qw      \\
\lstick{$\ket{\xi}$}    & \qw      & \swap{-1}  & \qw      & \qw
\end{quantikz}
\caption{}
\end{subfigure}
\begin{subfigure}[c]{0.49\textwidth}
\hspace*{-1.5cm}
\centering
\includegraphics[width = 1.2\textwidth]{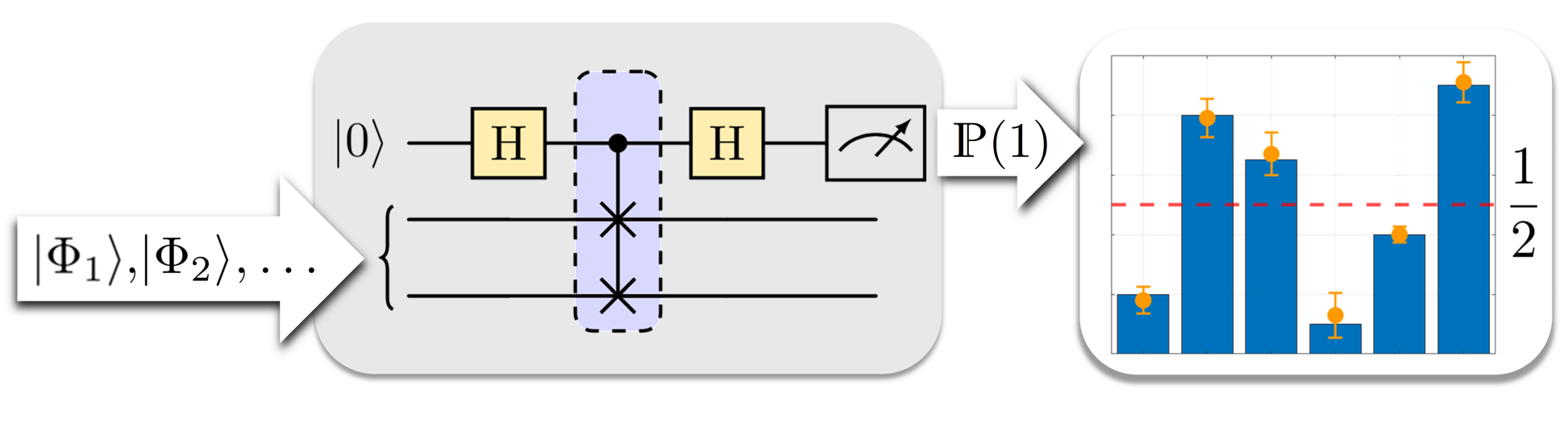}
\caption{}
\end{subfigure}

\caption{(a) Gate representation of the swap test algorithm performed on the states $\ket{\psi}$ and $\ket{\xi}$. The input qubits are initialised so that the first two are in the states $\ket{\psi}$ and $\ket{\xi}$ and the third one, which is an auxilliary qubit, is in the state $\ket{0}$. At the earth of the swap test there is a controlled-SWAP (CSWAP) gate, or Friedkin gate \cite{fredkin1982conservative}, that swaps the two target qubits only if the control qubit is in state $\ket{1}$. The Hadamard (H) gate \cite{nielsen2010quantum} is applied to the ancilla before the CSWAP gate and before the ancilla measurement. (b) The swap test circuit operating as an entanglement witness. For an arbitrary pure two-qubit input state, the probability of measuring the ancilla in $\ket{1}$ is extracted at the output. A value above $1/2$ guarantees that the state is entangled. The measured probability additionally yields a lower bound for the concurrence of the input state.}
\label{fig:SWAP_test_as_ent_witness}
\end{figure}
\subsection{$\mathds{P}(1)$ as an entanglement witness}
\label{sec:section_2.1}
The swap test \cite{barenco1997stabilization, buhrman2001quantum} is a quantum algorithm designed to assess the degree of overlap - quantified by the modulus square of the scalar product - between two unknown quantum states $\ket{\psi}$ and $\ket{\xi}$. The procedure is represented in terms of quantum gates in Fig. \ref{fig:SWAP_test_as_ent_witness}(a). The circuit translates the overlap estimation into a sampling task in which, by performing repeated measurements on an ancillary qubit, and analysing the outcome probability $\mathds{P}(1)$ (i.e., the probability of measuring the ancilla in state $\ket{1}$), one can efficiently compute the desired scalar product. More in detail, for an input state $\ket{\psi} \otimes \ket{\xi} \otimes \ket{0}$, where $\left\{\ket{\psi},\ket{\xi}\right\} \in \mathds{C}^2$ and the third qubit is the ancilla, the result reads as follows
\begin{equation}
    |\langle\psi | \xi\rangle|^2 = 1-2\mathds{P}(1)\,,
    \label{eq:swaptest_res}
\end{equation}
and trivially $\mathds{P}(0)=1-\mathds{P}(1)$.
Since $0 \leq |\langle \psi | \xi \rangle|^2 \leq 1$, then $0 \leq \mathds{P}(1) \leq 0.5$. So, for separable states, $\mathds{P}(1)$ and $\mathds{P}(0)$ are restricted to a subinterval of $[0,1]$ with a width of $1/2$.\\
Now, let us suppose that the target qubits are initially prepared in a generic pure two-qubit state $\ket{\Phi} \in \mathds{C}^2 \otimes \mathds{C}^2$. In the computational basis, this state can be decomposed as
\begin{equation}
    \ket{\Phi} = \alpha \ket{00} + \gamma\ket{10} + \delta \ket{01} + \beta \ket{11} \,,
    \label{eq:generic_two_qubit_state}
\end{equation}
where $\alpha, \beta, \gamma, \delta \in \mathds{C}$ with $|\alpha|^2 + |\beta|^2 + |\gamma|^2 + |\delta|^2 = 1$. When the swap test circuit is initially fed with the state $\ket{\Phi} \otimes \ket{0}$, the probability $\mathds{P}(1)_{\Phi}$ becomes 
\begin{align}
    \mathds{P}(1)_{\Phi} &= \mathrm{Tr}\left[\mathds{1}\otimes \ket{1}\bra{1} U_{\text{SWAP}}(\ket{\Phi}\bra{\Phi} \otimes \ket{0}\bra{0})U_{\text{SWAP}}^\dag\right] =\frac{1}{2}|\gamma-\delta|^2\,,\label{eq:P(1)_res}
\end{align}
where $\mathds{1}$ is the identity operator on $\mathds{C}^2 \otimes \mathds{C}^2$ and $U_{\text{SWAP}}$ is the unitary evolution implemented by the circuit. If the state is separable, the equation collapses to Eq.\eqref{eq:swaptest_res}. Generally, it is noteworthy that $\mathds{P}(1)_{\Phi}$ can now exceed the value of $1/2$. In particular, the following theorem holds.
\begin{theorem}
    Let $\ket{\Phi} \in \mathds{C}^2 \otimes \mathds{C}^2$ be a generic pure and normalized two-qubit state and let $\ket{\Phi} \otimes \ket{0}$ be the initial state provided as input to the circuit in Fig. \ref{fig:SWAP_test_as_ent_witness}(a). Then the condition
    \begin{equation}
    \mathds{P}(1)_{\Phi} > \frac{1}{2}
    \label{eq:witnessing_condition_pure}
\end{equation}
is a sufficient condition for $\ket{\Phi}$ to be entangled. 
\label{theo:ent_witness_pure}
\end{theorem}
\noindent A detailed proof of the theorem is provided in App.~\ref{app:A}.\\
For instance, let us consider the four Bell states
\begin{equation}
    \ket{\Phi^{\pm}} = \frac{1}{\sqrt{2}} \left( \ket{00} \pm \ket{11} \right)
    \quad,\quad
    \ket{\Psi^{\pm}} = \frac{1}{\sqrt{2}} \left( \ket{01} \pm \ket{10} \right) \,.
    \label{eq:bell_states}
\end{equation}
From Eq. \eqref{eq:P(1)_res}, we can note that only the state $\ket{\Psi^-}$ has unit $\mathds{P}(1)$, while the other three Bell states have $\mathds{P}(1)=0$. Thus, it is evident that the condition \eqref{eq:witnessing_condition_pure} is not necessary, but only sufficient.

The probability $\mathds{P}(1)$ also provides a direct way to estimate a lower bound on the concurrence $C(\Phi)$ \cite{wootters1998entanglement}. This is a particular entanglement monotone defined for bipartite two-qubit pure states of the form \eqref{eq:generic_two_qubit_state} as 
\begin{equation}
    C(\Phi):=2|\alpha\beta-\gamma\delta|\,.
\end{equation}It can be shown that $C(\Phi)$ attains values in the interval $[0,1]$. In particular $C(\Phi)=0$ if and only if $\ket{\Phi}$ is separable, while the value  $C(\Phi)=1$ corresponds to maximally entangled states. If $\mathds{P}(1)_\Phi\leq 1/2$, the bound is trivial ($C(\Phi)\geq 0$). For $\mathds{P}(1)_\Phi > 1/2$, let
\begin{equation}
    \epsilon := |\gamma - \delta|^2 - 1\,,
\end{equation}
quantifying the excess of $\mathds{P}(1)_\Phi$ above the separability threshold. By minimizing $C(\Phi)$ under normalization and fixed $\mathds{P}(1)_\Phi = (1+\epsilon)/2$ with $\epsilon \in (0,1]$, one finds 
\begin{equation}
    \min_{\mathds{P}(1)_\Phi = (1+\epsilon)/2} C(\Phi) = \epsilon \,.
\end{equation}
Hence, the relationship between $\mathds{P}(1)_\Phi$ and the concurrence can be formally stated as follows.
\begin{theorem}\label{theo:conc_bound}
Let $\ket{\Phi} \in \mathds{C}^2 \otimes \mathds{C}^2$ be a generic pure and normalized two-qubit state, and let $\mathds{P}(1)_\Phi$ denote the probability of obtaining outcome $1$ from the swap test circuit in Fig. \ref{fig:SWAP_test_as_ent_witness}(a). 
Then the concurrence $C(\Phi)$ satisfies the lower bound
\begin{equation}\label{FinalBoundConcurrencePureGeneral}
    C(\Phi) \ge f\big(\mathds{P}(1)_\Phi\big)\,,
\end{equation}
where the function $f:[0,1] \to \mathds{R}$ is defined as
\begin{equation}
f(x) =
\begin{cases}
0\,, & x \le \tfrac{1}{2}\,,\\[6pt]
2x - 1\,, & x > \tfrac{1}{2}\,.
\end{cases}
\label{eq:function_bound}
\end{equation}
\end{theorem}
\noindent A full derivation is provided in App.~\ref{app:low_bound_conc_pure}.

These theorems state that $\mathds{P}(1)_\Phi$ can be taken as an observable for entanglement witnessing and its value provides a lower bound for the concurrence $C(\Phi)$ of a generic two-qubit pure state $\ket{\Phi}$, Eq. \eqref{eq:generic_two_qubit_state}. The working principle of the protocol is illustrated in Fig. \ref{fig:SWAP_test_as_ent_witness}(b): an arbitrary two-qubit state is injected into the swap test circuit, and the probability of observing the ancilla in $\ket{1}$ is obtained at the output. When this probability exceeds $1/2$, the input state is certified to be entangled, and its value also provides a lower bound on the state concurrence.
\subsection{Extension to mixed states}
\label{sec:section_2.2}
In experimental implementations of quantum algorithms, pure quantum states are rarely preserved due to unavoidable interactions with the environment and imperfections in real quantum hardware. As a result of this loss of coherence and purity, systems are typically described by mixed states. To analyze the robustness of the swap test for entanglement witnessing, let us consider the generic case of a mixed state $\rho$ of a bipartite two-qubit system. As in Sec.~\ref{sec:section_2.1}, we shall denote with $\mathds{P}(1)_\rho$ the probability of detecting the state $\ket{1}$ of the ancilla qubit if the swap test circuit is initially fed with the state $\rho \otimes \ket{0}$.
More specifically, we have
\begin{equation}
\mathds{P}(1)_\rho=\mathrm{Tr}\left[\mathds{1}\otimes |1\rangle\langle 1|U_{\text{SWAP}}(\rho\otimes |0\rangle\langle 0|)U_{\text{SWAP}}^\dag\right]\,.
\end{equation}
The concurrence of a general two-qubit mixed state $\rho$ is the entanglement measure defined as \cite{wootters1998entanglement, mintert2005measures}
\begin{equation}
    C(\rho):=\min_{\{p_i, |\psi_i\rangle\}}\sum _ip_iC( |\psi_i\rangle)\,,
\end{equation}
where the  minimization is over all pure-state decompositions of $\rho$:
\begin{equation}
    \rho=\sum_ip_i|\psi_i\rangle\langle \psi_i|\,,\quad
    \mbox{with}\,\,p_i\geq 0\,, \, \sum _ip_i=1\,.
    \label{eq:mixed_states}
\end{equation}
As in the case of pure states, $C(\rho)\in [0,1]$ and $C(\rho)=0$ if and only if $\rho$ is separable. In addition, the bound in Eq. \eqref{FinalBoundConcurrencePureGeneral} can be extended to general mixed states. Indeed for any decomposition  $\rho=\sum_ip_i|\psi_i\rangle\langle \psi_i|$ we have 
\begin{align}
    \sum _ip_iC( |\psi_i\rangle)&\geq \sum _ip_i f(\mathds{P}(1)_{\psi_i}) \nonumber\\
    &\geq f\left(\sum _ip_i \mathds{P}(1)_{\psi_i}\right)
    =f(\mathds{P}(1)_\rho) \,,
\end{align}
where the first inequality is a consequence of Eq. \eqref{FinalBoundConcurrencePureGeneral}, the second one is due to the convexity of the map $f$ and the last equality follows by the linearity of $\mathds{P}(1)_\rho$. Hence, by the values attained by the function $f$, we can conclude that if $\mathds{P}(1)_\rho > 1/2$ then the state $\rho$ is entangled.

The specific form of mixed state depends in general on the details of the experiment and on the noise processes at play. An interesting class of mixed states is represented by Werner states \cite{werner1989quantum}. Specifically, let us consider the family of Werner-like mixed states of the form
\begin{equation} \label{eq:Werner-state}
    \rho = p \ket{\Phi}\bra{\Phi} + \frac{1-p}{4}\mathds{1}\,,
\end{equation}
where $\ket{\Phi}$ is the generic pure two-qubit state introduced in \eqref{eq:generic_two_qubit_state} and $p \in [0,1]$. 
Proper Werner states are recovered when $\ket{\Phi}$ is one of the Bell states, Eq.\eqref{eq:bell_states}. The probability of obtaining the state $\ket{1}$ for the auxiliary qubit in the swap test circuit initially fed with the state \eqref{eq:Werner-state} is 
\begin{align}
    \mathds{P}(1)_\rho& =\mathrm{Tr} \left[\mathds{1}\otimes |1\rangle\langle 1|U_{\text{SWAP}}(\rho\otimes|0\rangle\langle 0|)U_{\text{SWAP}}\right]\nonumber\\
    &=p\frac{|\gamma-\delta|^2}{2} +\frac{(1-p)}{4}\label{eq:P1-werner}\,.
\end{align}
For instance, when $\ket{\Phi}$ correspond to the Bell state $\ket{\Psi^-}$ the probability reads 
\begin{equation}
    \mathds{P}(1)_\rho = \frac{3}{4}p + \frac{1}{4}\,,
\end{equation}
while for the other Bell states is equal to $(1-p)/4$.
Exploiting the PPT criterion \cite{peres1996separability, Horodecki:1996nc} it is possible to show that $\rho$ is separable if and only if the condition
\begin{equation}
    p \leq \frac{1}{1+4|\alpha \beta - \gamma \delta|}  = \frac{1}{1+2C(\Phi)}
\end{equation}
is satisfied, and that $\mathds{P}(1)_\rho > 1/2$ is a sufficient condition for $\rho$ to be entangled, in agreement with the general case. Further details are provided in App.~\ref{app:werner_states}.
\subsection{Enhancing detection via local unitary pre-processing}
\label{sec:sec_enhancing}
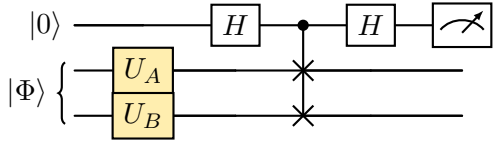
\begin{figure}
    \centering
        \begin{quantikz}[row sep={0.6cm,between origins}, column sep=0.5cm]
        \lstick{$\ket{0}$} & \qw      & \gate{H} & \ctrl{2} & \gate{H} & \meter{} \\
        \lstick[2]{$\ket{\Phi}$} &\gate[style={fill=yellow_sdrac}]{U_A}  & \qw      & \targX{}   & \qw      & \qw      \\
        & \gate[style={fill=yellow_sdrac}]{U_B}  & \qw & \swap{-1}  & \qw      & \qw
        \end{quantikz}
    \caption{Modified swap test circuit with a reconfigurable local unitary $U = U_A \otimes U_B$ applied to the input state. The circuit is run multiple times with different settings of $U$ to enhance entanglement detection.}
    \label{fig:SWAP_modified}
\end{figure}
The proposed entanglement witness is selective: numerical simulations\footnote{In these simulations, the swap test was implemented using \texttt{Qiskit}, and entanglement was evaluated by computing the Schmidt rank of each state.} show that approximately $12.5\%$ of random entangled pure states (sampled from the unit sphere in $\mathds{C}^4$) satisfy the detection criterion $\mathds{P}(1) > 1/2$. This limitation can be mitigated by introducing a simple local pre-processing stage prior to the swap test, as shown in Fig. \ref{fig:SWAP_modified}. In this scheme, $U = U_A \otimes U_B$ is a reconfigurable local unitary transformation applied to the two-qubit input state $\ket{\Phi}$. The circuit is executed four times, each with a different configuration of $U$:
\begin{align}
    U &= U_1 = \mathds{1}\otimes\mathds{1} \quad (\text{standard swap test circuit})\,; \\
    U &= U_2 = \mathds{1}\otimes\sigma_z\,; \\
    U &= U_3 = \mathds{1}\otimes\sigma_x\,; \\
    U &= U_4 = \sigma_x \otimes\sigma_z\,,
\end{align}
where $\sigma_i$ ($i = \{x,y,z\}$) are the Pauli matrices. These specific unitaries are chosen because they map the three Bell states $\ket{\Psi^+}$, $\ket{\Phi^-}$, and $\ket{\Phi^+}$ into $\ket{\Psi^-}$ up to a global phase. Importantly, all these transformations are local and therefore preserve the entanglement content of the state $\ket{\Phi}$. Consequently, if any of the four corresponding measurements of $\mathds{P}(1)$ satisfies $\mathds{P}(1) > 1/2$, one can safely conclude that $\ket{\Phi}$ is entangled. Numerical simulations confirm that this approach significantly increases the entanglement detection rate from $12.5\%$ in the standard swap test to about $50\%$ using the modified protocol.
\subsection{Quantum random number generation}
The entanglement witness characteristic of the swap test can also be exploited in the certification of the min-entropy \cite{konig2009operational} of a Quantum Random Number Generator (QRNG) \cite{herrero2017quantum}. In the specific case of a pure state $\ket{\Phi}$ described by Eq. \eqref{eq:generic_two_qubit_state} , if $\mathds{P}(1)_{\Phi} > 1/2 $ then the lower bound on the concurrence given by Theorem \ref{theo:conc_bound} yields an upper bound for the quantum guessing probability $G(\Phi)$ of the form
\begin{equation}
    G(\Phi) \leq \frac{1+\sqrt{1-f^2(\mathds{P}(1)_{\Phi})}}{2}\,,
\end{equation}
where $f: [0,1] \rightarrow \mathds{R}$ is defined by Eq. \eqref{eq:function_bound}, and from $G(\Phi)$ one can get the min-entropy $H_{\text{min}}(\Phi) = - \log_2\left(G(\Phi)\right)$. In presence of non-idealities, Eq. \eqref{eq:function_bound_non_ideal} must be considered instead. This result can also be generalized to arbitrary mixed states.  More details are provided in Sec. \ref{sec:section_3.2} and App.~\ref{app:qu_randomness}.
\section{Integrated photonic implementation of the SWAP-based entanglement witness}
\label{sec:section_3}
\subsection{Linear photonic swap test circuit}
\label{sec:section_3.1}
\begin{figure}
\centering
\begin{subfigure}[t]{0.49\textwidth}
\centering
\begin{quantikz}[row sep={0.6cm,between origins}, column sep=0.5cm]
\lstick{$\ket{0}$}      & \gate[style={fill=yellow_sdrac}]{\text{MMI}} & \ctrl{2}\gategroup[3,steps=1,style={dashed,rounded
 corners,fill=blue!15, inner
 xsep=4pt},background,label style={label
 position=below,anchor=north,yshift=-0.2cm}]{CSWAP}  & \gate[3]{\text{PS}_3} & \gate[style={fill=yellow_sdrac}]{\text{MMI}} & \meter{} \\
\lstick{$\ket{\psi}$}   & \qw      & \targX{} & \qw & \qw      & \qw      \\
\lstick{$\ket{\xi}$}    & \qw      & \swap{-1} & \qw & \qw      & \qw
\end{quantikz}
\caption{}
\end{subfigure}
\begin{subfigure}[t]{1\textwidth}
\centering
\includegraphics[width=\textwidth]{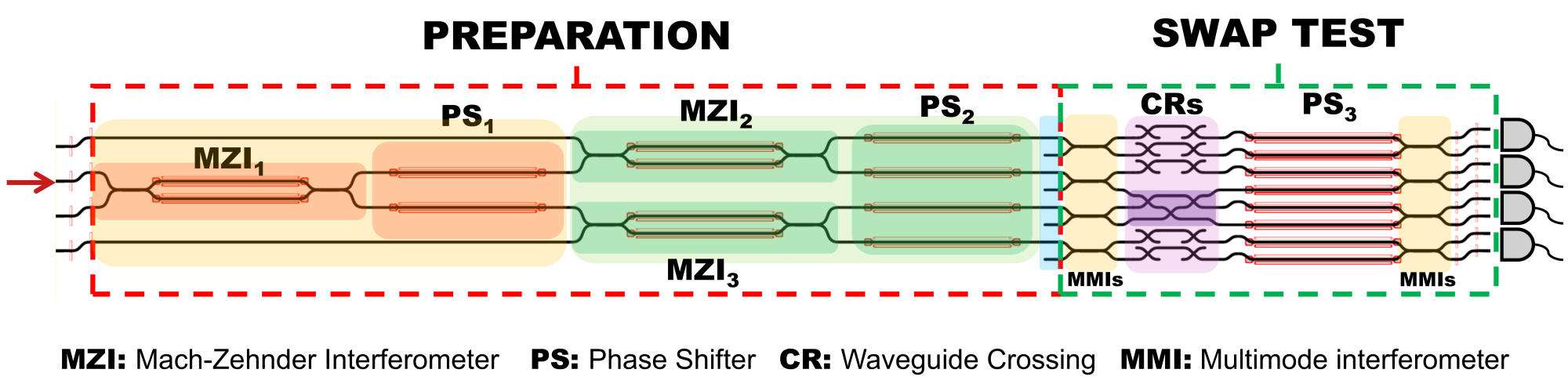}
\caption{}
\end{subfigure}

\caption{(a) Schematic gate representation of the linear photonic implementation of the swap test circuit. The H gates are implemented through MMIs, i.e. integrated beam-splitters, and the gate $\text{PS}_3$ represents the action of PSs needed to counterbalance spurious phases between the optical paths. (b) Design of the PIC that implements the swap test circuit \cite{baldazzi2024linear}. Waveguides are depicted in black while the red stripes identify the sections where thermal phase shifters are present. The red arrow indicates the waveguide where light is injected. The red box contains the preparation stage, where integrated MZIs and PSs are used to encode the input state $\ket{\Phi} \otimes \ket{0}$, Eq. \eqref{eq:generic_two_qubit_state}. The green box contains the swap stage where MMIs, CRs and PSs are exploited to implement the swap test algorithm. Referring to (a), we see the sequence of MMIs (orange), the CSWAP gate (violet), the $\text{PS}_3$ gate (white), and the second set of MMIs (orange). Finally, single photons are detected at the output.}
\label{fig:SWAP_test_photonic_gate_repr}
\end{figure}
Among the various hardware platforms available for quantum technologies, integrated photonics has emerged as a particularly promising candidate for implementing low-noise quantum information processing \cite{slussarenko2019photonic, wang2020integrated,romero2024photonic}. Photons are natural and excellent carriers of quantum information, as they interact weakly with the environment and can be linearly manipulated relatively easily at room temperature. Moreover, thanks to advances in nanofabrication, it is now possible to implement complex quantum circuits directly on-chip, using reconfigurable networks of well-established optical components such as beam splitters, phase shifters, and interferometers \cite{giordani2023integrated}. Specifically, the swap test algorithm can be implemented on a photonic integrated circuit \cite{baldazzi2024linear} relying solely on linear optical integrated components and qudits, represented by single photons from an attenuated laser beam propagating through a set of waveguides. The circuit design is presented in Fig. \ref{fig:SWAP_test_photonic_gate_repr}(b). \\
The PIC is divided into two sections: the first one is the preparation stage, and the second one is the swap test stage. To encode states in the PIC, path-encoded single photons are used, where the state is determined by the waveguide through which a single photon travels. For instance, in a four-waveguide system containing a single photon, we can identify the four computational basis states $\ket{00}$, $\ket{01}$, $\ket{10}$ and $\ket{11}$ with the photon propagating in the first, second, third, and fourth waveguide, respectively. Further details are provided in App.~\ref{app:analytical_description}. The manipulation of the quantum states is achieved by three integrated Mach-Zehnder interferometers (MZIs) and six phase shifters (PSs), divided into two layers. The action of this triangular arrangement of MZIs and PSs on the input photon state $\ket{01}$ allows the preparation of any two-qubit state $\ket{\Phi}$, including entangled states. In the final section of the preparation stage, highlighted within the blue box in Fig. \ref{fig:SWAP_test_photonic_gate_repr}(b), four additional waveguides are introduced and connecting the outputs of $\text{PS}_2$ to the even-numbered waveguides (with waveguides indexed starting from $0$) passively initialises the auxiliary qubit to the state $\ket{0}$. Therefore, the total state at the end of the preparation stage reads $\ket{\Psi} = \ket{\Phi} \otimes \ket{0}$ where $\ket{\Phi}$ is the generic two-qubit state, Eq. \eqref{eq:generic_two_qubit_state}, produced by the action of the nested configuration of MZIs and PSs. Specifically, let \( \boldsymbol{\theta}_1 = (\theta_1(1), \theta_1(2)), \boldsymbol{\phi}_1 = (\phi_1(1), \phi_1(2)) \) denote the phases of the two arms of \(\text{MZI}_1\) and of \(\text{PS}_1\), respectively. Similarly, let \( \boldsymbol{\theta}_{21} = (\theta_{21}(1), \theta_{21}(2)), \boldsymbol{\phi}_{21} = (\phi_{21}(1), \phi_{21}(2)) \) denote the phases of \(\text{MZI}_2\) and the first pair of phase shifters in \(\text{PS}_2\), and let \( \boldsymbol{\theta}_{22} = (\theta_{22}(1), \theta_{22}(2)), \boldsymbol{\phi}_{22} = (\phi_{22}(1), \phi_{22}(2)) \) denote the phases of \(\text{MZI}_3\) and the second pair of phase shifters in \(\text{PS}_2\). Then, the state  $\ket{\Phi}$ can be written - up to a global phase - in the same form of Eq. \eqref{eq:generic_two_qubit_state} with
\begin{equation}
    \begin{split}
       \alpha &= {\rm e}^{\text{i} (\phi_1(1)-(\theta_{21}(1)+\theta_{21}(2)) + \phi_{21}(1))} \sin \Delta \theta_1 \cos \Delta \theta_{21}\,,\\
       \gamma &= {\rm e}^{\text{i} (\phi_1(2)-(\theta_{22}(1)+\theta_{22}(2)) + \phi_{22}(1))} \cos \Delta \theta_1 \sin \Delta \theta_{22}\,,\\
       \delta &= {\rm e}^{\text{i} (\phi_1(1)-(\theta_{21}(1)+\theta_{21}(2)) + \phi_{21}(1))} {\rm e}^{-\text{i}\Delta \phi_{21}} \sin \Delta \theta_1 \sin \Delta \theta_{21}\,,\\
       \beta &= {\rm e}^{\text{i} (\phi_1(2)-(\theta_{22}(1)+\theta_{22}(2)) + \phi_{22}(1))} {\rm e}^{-\text{i}\Delta \phi_{22}} \cos \Delta \theta_1 \cos \Delta \theta_{22}\,,
    \end{split}
    \label{eq:generic_prepared_state}
\end{equation}
where $\Delta \phi_j \equiv \phi_j(1) - \phi_j(2)$. The detailed derivation of Eq. \eqref{eq:generic_prepared_state} is provided in App. \ref{app:analytical_description}.\\
The swap stage of the circuit implements the algorithm through a series of multimode interferometers (MMIs), PSs, and a network of waveguide crossings (CRs). The MMIs are integrated beam splitters and play the role of Hadamard gates on the auxiliary qubit, with the only difference that $U_{\text{MMI}}^2$ is not equal to the identity but to ${\rm i}\sigma_x$, leading to an exchange of the outcome probabilities $\mathds{P}(1)$ and $\mathds{P}(0)$ compared to the standard swap test. The CSWAP operation is realized through three cascaded waveguide crossings, while an array of tunable phase shifters compensates for path-dependent phase errors introduced during fabrication. The overall unitary describing the swap test stage is
\begin{equation}
U_{\text{swaptest}}(\boldsymbol{\theta}s) = (\mathds{1} \otimes U_{\text{MMI}}) \cdot U_{\text{PS}3}(\boldsymbol{\theta}s) \cdot U_{\text{CSWAP}}\cdot (\mathds{1} \otimes U_{\text{MMI}})\,,
\end{equation}
which reproduces the expected swap test probability in Eq. \eqref{eq:P(1)_res} with $\mathds{P}(0)$ playing the role of $\mathds{P}(1)$. Hence, the photonic integrated circuit enables the successful implementation of the swap test algorithm. 
From an experimental point of view, the probabilities $\mathds{P}(0)$ and $\mathds{P}(1)$ are estimated by sampling the photon counts at the outputs of the circuit. To avoid any ambiguity arising from the specific behaviour of our photonic implementation and, thus, to obtain the same outcomes of the original algorithm described in Sec.~\ref{sec:section_2.1}, they are directly computed as
\begin{equation}
    \mathds{P}(x) = \frac{N_{1-x}}{N_0 + N_1} \quad (x = 0,1)\,,
    \label{eq:prob_to_freq}
\end{equation}
where $N_0$ and $N_1$ are the total counts coming from waveguides corresponding to the auxiliary qubit in state $\ket{0}$ and $\ket{1}$ respectively.
For a more detailed description of the circuit refer to App.~\ref{app:analytical_description} and \cite{baldazzi2024linear}.

The photonic implementation of the swap test circuit offers several advantages. Primarily, the use of path encoding enables a fully linear, on chip optical circuit, relying only on well-established components. This choice results in a short and simple circuit. Another key benefit of having a linear circuit is the ability to use an attenuated laser source in place of true single-photon sources. Indeed, this is possible because all operations in the circuit are linear at the single-photon level \cite{baldazzi2024linear,pasini2020bell}. In addition, the PIC operates without the need for cryogenic temperatures or complex optical setups. Furthermore, the use of linear photonic integrated circuits allows for the experimental implementation of the reconfigurable local unitary $U$ introduced in Sec.~\ref{sec:sec_enhancing} to enhance the detection capability of the witness. This can be realized, for example, using universal unitary decompositions such as the Reck or Clements schemes \cite{reck1994experimental,clements2016optimal}.

\subsection{Robustness to experimental noise and imperfections}
\label{sec:section_3.2}
In Sec. \ref{sec:section_2.2} Theorems \ref{theo:ent_witness_pure} and \ref{theo:conc_bound} have been extended to mixed states, demonstrating that the swap test can also be used to witness entanglement in a more general framework. However, real platform-specific imperfections introduce additional noise. In our photonic chip, non-ideal optical components affect not only the state purity but also the measured values of $\mathds{P}(1)$, thus shifting the separability threshold. In particular, we focus on the following non-idealities \cite{baldazzi2024linear}
\begin{itemize}
    \item Non-ideal MMIs: due to fabrication errors, the multimode interferometers are not exactly implementing the matrix of an ideal $50:50$ beam splitter. The transmission $t$ and reflection $r$ coefficients might be different from $1/\sqrt{2}$. At the experimental operating wavelength of $750$ nm, the coefficients were measured to be $t^2 = 0.48(2)$ and $r^2 = 0.52(2)$.
    \item Phase errors: calibration inaccuracies, power supply errors together with current and temperature fluctuations lead to deviations from the exact values of the phase to which every PS is nominally set. Relative phase errors are assumed to be random variables normally distributed with mean $0$ and standard deviation $\sigma = 0.1$ rad.  
\end{itemize}
Their effect is to modify the maximum value of $\mathds{P}(1)$ over the set of separable states. Assuming\footnote{In presence of losses, $t^2 + r^2 < 1$, which can be expressed as $t^2 + r^2 = {\rm e}^{\alpha_{\rm loss}}$ with $\alpha_{\rm loss} < 0$. However, under the assumption of identical MMIs' losses, the measured probabilities defined by Eq. \eqref{eq:prob_to_freq} remain unchanged with respect to the lossless case, as they are normalized to the total number of detected photons.} $t^2 + r^2 = 1$, the separability threshold shifts from $1/2$ to 
\begin{equation}
\label{eq:non_ideal_threshold}
    \frac{1}{2} + \frac{c}{2}\,,
\end{equation}
where $c = t^4 + r^4 - 2r^2t^2\cos{\sigma}$. For the measured values $t^2 = 0.48$, $r^2 = 0.52$ and $\sigma = 0.1$, we find $c = 0.004$. To assess the worst case scenario, we consider a $2 \sigma$ confidence interval for both $t^2$ and $r^2$, i.e., $t^2 \in  [0.44,0.52]$ and $r^2 \in [0.48, 0.56]$, while maintaining $t^2 + r^2 = 1$. The maximum deviation occurs for the largest imbalance, $t^2 = 0.44$ and $r^2 = 0.56$, yielding $c \approx 0.017$. Accordingly, in the certification of entanglement, we consider Eq. \eqref{eq:non_ideal_threshold} with this value of $c$ as the effective separability threshold, accounting for typical phase errors and approximately $95\%$ confidence bounds on the MMI's transmittivities and reflectivities. The small relative variation of the threshold demonstrates the robustness of the entanglement witness against realistic circuit imperfections.\\
Then, let us assume that $\mathds{P}(1)=\frac{c+1+\epsilon}{2}$, with $0<\epsilon<2t^2r^2(1+\cos\sigma)$. In this case the state is entangled and its concurrence is strictly positive. In particular, the concurrence of a pure input state satisfies
\begin{equation}
C(\Phi) \geq \tilde{f}(\mathds{P}(1))\,
\end{equation}
with $\tilde{f}: [0,1] \rightarrow \mathds {R}$ defined as
\begin{equation}
\tilde{f}(x) =
\begin{cases}
0\,, & 0 \le x \le \tfrac{1+c}{2}\,,\\[6pt]
\frac{2x-1-c}{2t^2r^2(1+\cos\sigma)}\,, & \tfrac{1+c}{2}<x\le 1\,.
\end{cases}
\label{eq:function_bound_non_ideal}
\end{equation}
As expected, the minimum of the threshold $\mathds{P}(1)=\frac{c+1}{2}$ and the maximum of the concurrence lower bound are obtained for $\sigma=0$ and balanced MMIs, i.e. $t^2 = r^2 = 1/2$. A detailed derivation of the results presented is provided in App.~\ref{app:non_idealities}.
\section{Results}
\label{sec:section_4}
\begin{figure}
    \centering
    \begin{subfigure}{0.49\textwidth}
        \centering
         \includegraphics[width = \textwidth]{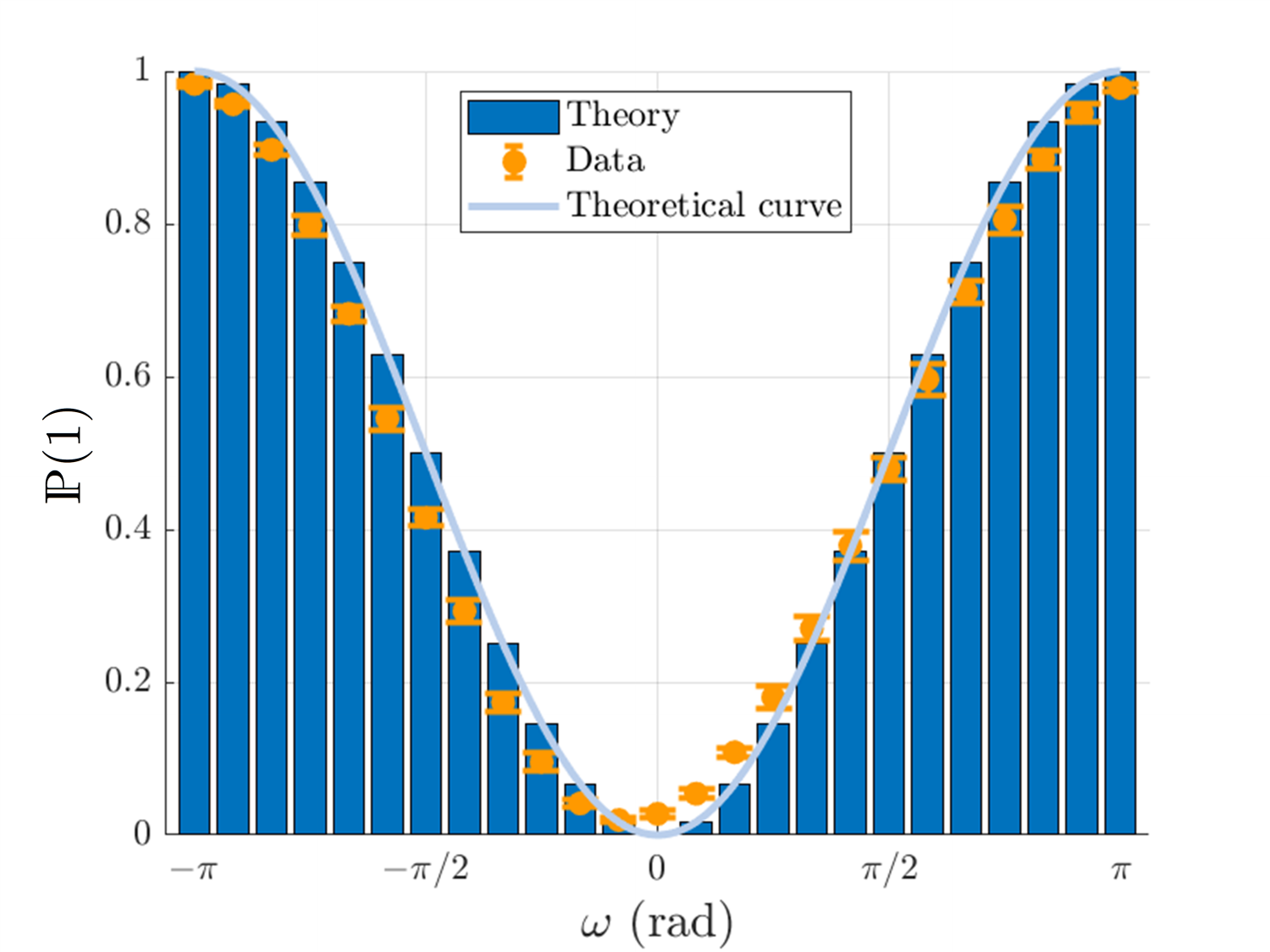}
         \caption{}
    \end{subfigure}
    \begin{subfigure}{0.49\textwidth}
        \centering
         \includegraphics[width = \textwidth]{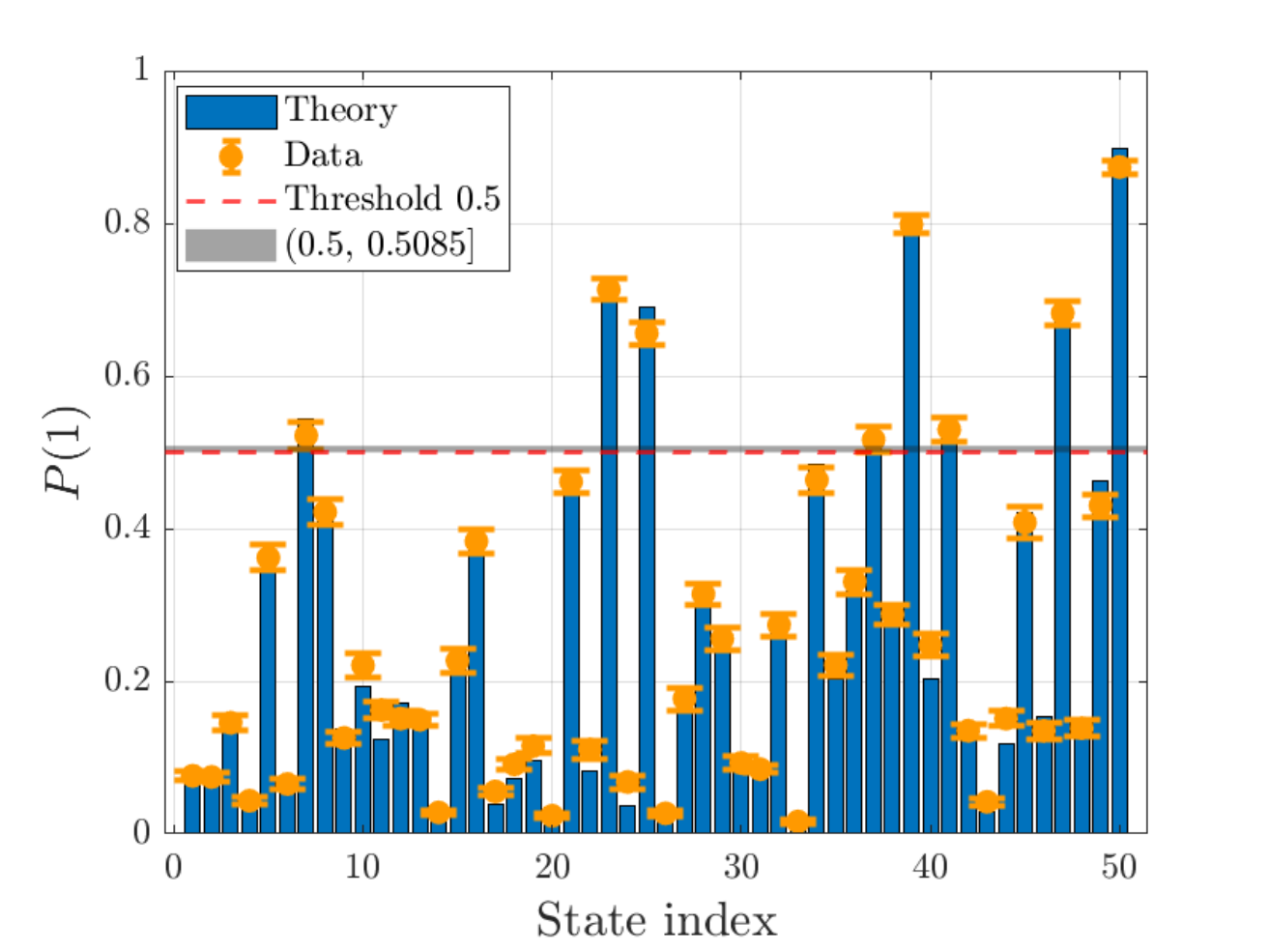}
         \caption{}
    \end{subfigure}
    \begin{subfigure}{0.49\textwidth}
        \centering
         \includegraphics[width = \textwidth]{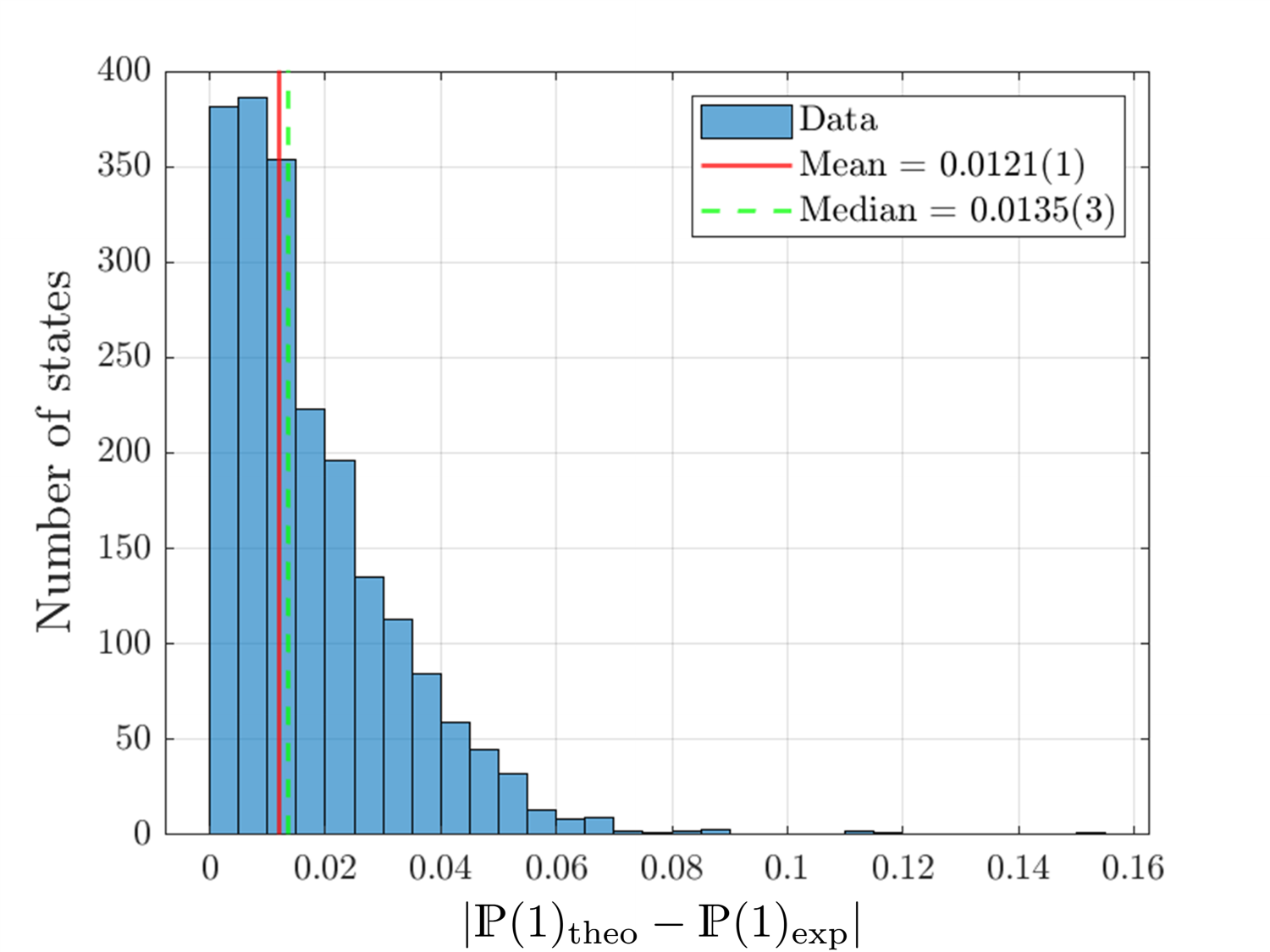}
         \caption{}
    \end{subfigure}
    \begin{subfigure}{0.49\textwidth}
        \centering
         \includegraphics[width = \textwidth]{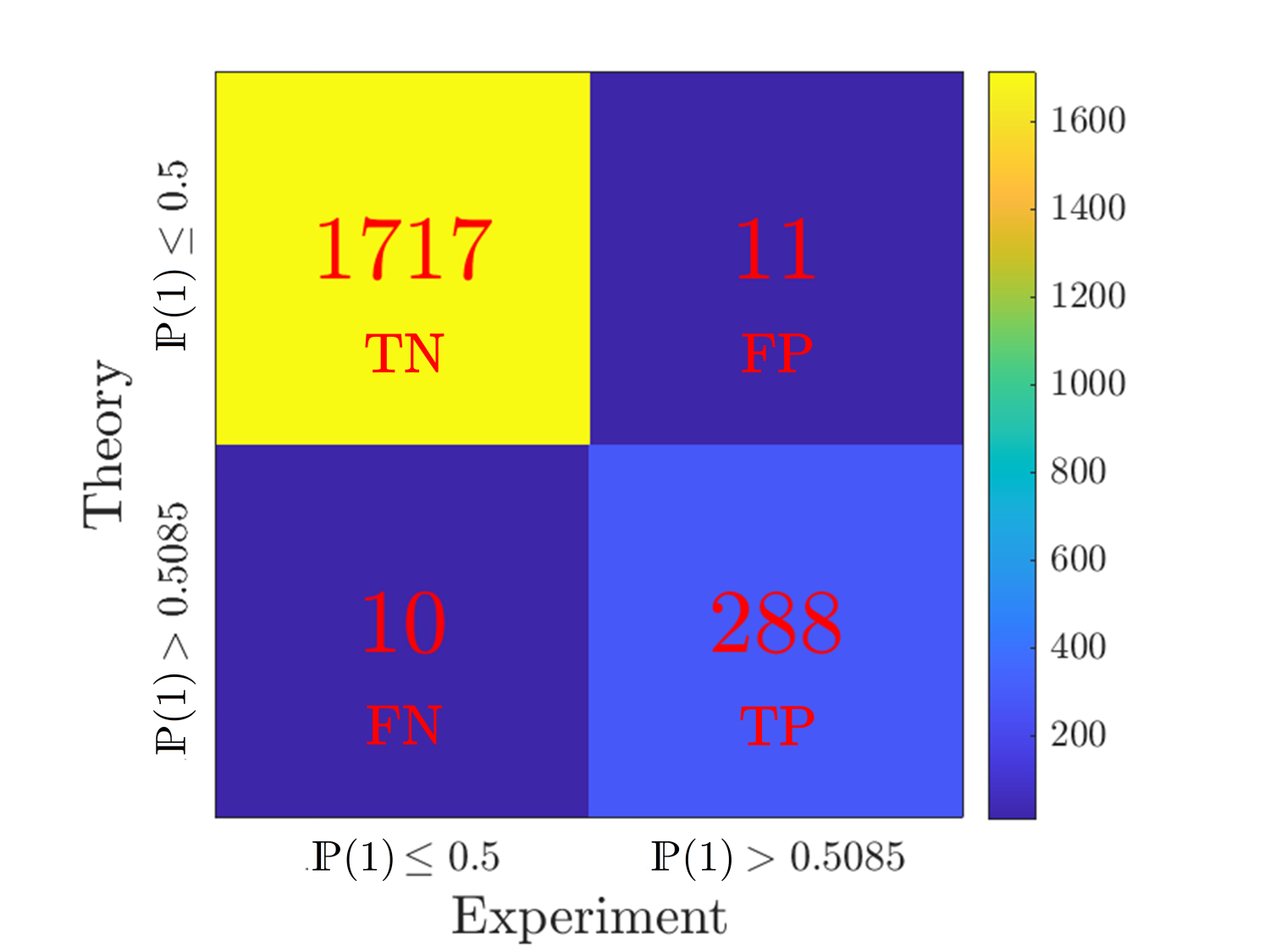}
         \caption{}
    \end{subfigure}
    
    \caption{(a) Experimental results for the family of states described by Eq. \eqref{eq:psi_omega}. Theoretical values of $\mathds{P}(1)$ are represented by blue bars connected by a light blue curve, while experimental data with error bars are shown in orange. (b) Subset of $50$ two-qubit states randomly selected from a total of $2050$. Theoretical values of $\mathds{P}(1)$ are shown in blue, and experimental results with error bars in orange. The dashed red line at $\mathds{P}(1) = 0.5$ indicates the threshold for entanglement witnessing of pure states, as established in Theorem \ref{theo:ent_witness_pure}. The shaded grey region between $0.5$ and $0.5085$ represents the range of $\mathds{P}(1)$ values where entanglement cannot be confidently detected due to non-idealities in the PIC (see Sec.~\ref{sec:section_3.2}).  (c) Histogram of the distance in modulus between the theoretical and experimental results for $\mathds{P}(1)$. The vertical lines indicate the mean and median. The width of the bins, computed using Scott's rule, is $0.005$, and the bins are centered around multiple values of $0.0025$. (d) Confusion matrix summarizing the performance of the circuit as a classifier. States with a theoretical $\mathds{P}(1)\in [0.5,0.5085]$ are excluded from the analysis since, even if the measured $\mathds{P}(1)$ perfectly matches the theory: entanglement cannot be confidently identified for these cases. The overall accuracy is $99.0(1)\%$. }
    \label{fig:random_results}
\end{figure} 
Using the same PIC presented in \cite{baldazzi2024linear}, we validate the swap test and its photonic implementation as entanglement identifier and quantifier for two-qubit pure states, Eq. \eqref{eq:generic_two_qubit_state}, and Werner-like mixed states, Eq. \eqref{eq:Werner-state}. The experimental procedures for data acquisition are described in detail in the Methods section.

As a first set of entangled states we consider the four Bell states, Eq. \eqref{eq:bell_states}. 
The experimental results and theoretical predictions are summarized in Tab. \ref{tab:results_bell}. The states $\ket{\Phi^+}$ and $\ket{\Phi^-}$ are grouped together as the PIC output does not provide access to the relative phase between $\ket{00}$ and $\ket{11}$. The deviations of the measured values from the ideal expectations arise from circuit non-idealities. When these effects are taken into account, the measurement values agree with the theoretical predictions corrected with non-idealities within $2 \sigma$. Additional details are provided in App.~\ref{app:non_idealities_2sigma}.
\begin{table}
\renewcommand{\arraystretch}{1.5}
\centering
\caption{
Theoretical predictions and experimental results for the values of $\mathds{P}(1)$ for the two-qubit Bell states defined in Eq. \eqref{eq:bell_states}. $\ket{\Phi^+}$ and $\ket{\Phi^-}$ are grouped together since the PIC output is insensitive to the relative phase between $\ket{00}$ and $\ket{11}$. }
\label{tab:results_bell}
\begin{tabular}{ c c c c }
\toprule
\ &  $\ket{\Phi^\pm}$ & $\ket{\Psi^+}$ &  $\ket{\Psi^-}$ \\
\midrule
Theoretical $\mathds{P}(1)$ & 0 & 0 & 1   \\
Experimental $\mathds{P}(1)$ & 0.013(2) & 0.016(4) & 0.98(2)   \\
\bottomrule
\end{tabular}
\end{table}

The second set of states is given by the following one-parameter family of states:
\begin{equation}
    \ket{\Psi(\omega)} = \frac{1}{\sqrt{2}} \left( \ket{01} + {\rm e}^{{\rm i}\omega} \ket{10} \right) \,.
    \label{eq:psi_omega}
\end{equation}
These states interpolate between $\ket{\Psi^+}$ and $\ket{\Psi^-}$, and the result of swap test is simply $\mathds{P}(1)=\sin^2 \left( \omega/2 \right)$. Fig. \ref{fig:random_results}(a) shows the experimental and theoretical data for different values of $\omega$ in the interval $[-\pi,\pi]$, where the borders of the interval correspond to $\ket{\Psi^-}$. The observed root mean square error (RMSE) is 0.04.

As third validation step, we consider the preparation of generic random two-qubit states by selecting a 2050 random set of values for the phases $\boldsymbol{\theta}$s and $\boldsymbol{\phi}$s in Eq.\eqref{eq:generic_prepared_state}. This procedure aims to test the entanglement witness described in Sec.~\ref{sec:section_2.1} with a generic two-qubit state, Eq.\eqref{eq:generic_two_qubit_state}. Since the condition for the entanglement witness is only sufficient, a positive result of the algorithm, i.e. $\mathds{P}(1)>0.5085$, means that the state is entangled, while no conclusion can be drawn from a negative result. Thus, the PIC can recognize only a subset of the whole space of entangled two-qubit states.
Fig. \ref{fig:random_results}(b) shows a subset of 50 states randomly chosen from the 2050 tested states. The shaded grey region in Fig. \ref{fig:random_results}(b) highlights states characterized by $\mathds{P}(1)\in [0.5,0.5085]$. The entanglement of these states cannot be confidently witnessed because of the non-idealities in the PIC, as discussed in Sec.~\ref{sec:section_3.2}. Among the 2050 tested states, 24 yield $\mathds{P}(1)\in[0.5,0.5085]$ and are therefore excluded from the comparison between theory and experiments, even if their measured $\mathds{P}(1)$ is perfectly compatible with the theory.
We quantified the quality of our estimation of $\mathds{P}(1)$ for the random 2050 states by using the distance between the theoretical and experimental $\mathds{P}(1)$, i.e., $\mathds{P}(1)_{\text{theo}}-\mathds{P}(1)_{\text{exp}}$. All distance values between theory and experiments lie within the range $[-0.1139(9),0.15(1)]$, the data have weighted mean and median equal to $-0.0062(1)$ and $-0.0047(5)$ respectively, and the skewness of the distribution is $0.13$. The asymmetry is related to the observed tendency of the PIC to slightly overestimate the value of $\mathds{P}(1)$. The histogram in Fig. \ref{fig:random_results}(c) displays the distance modulus, i.e. $|\mathds{P}(1)_{\text{theo}}-\mathds{P}(1)_{\text{exp}}|$. 
The mean and the median of the distance moduli over the 2050 tested states are respectively $0.0121(1)$ and $0.0135(3)$, and the observed RMSE for the random states is $0.023$.
In Fig. \ref{fig:random_results}(d), we summarize the result with the random states in the confusion matrix, where we have true positives (\textbf{TP}) and true negatives (\textbf{TN}) as diagonal elements and false negatives (\textbf{FN}) and false positives (\textbf{FP}) as off-diagonal elements. In our case, \textbf{TN}/\textbf{TP} is the number of states for which $\mathds{P}(1) \le 0.5$/$\mathds{P}(1) > 0.5085$ holds both theoretically and experimentally, \textbf{FP} the number of states with $\mathds{P}(1)_{\text{theo}} \le 0.5$ and $\mathds{P}(1)_{\text{exp}} > 0.5085$, and \textbf{FN} the number of states with $\mathds{P}(1)_{\text{theo}} > 0.5085$ and $\mathds{P}(1)_{\text{exp}} \le 0.5$. More explicitly, \textbf{FP}s are the states recognized as entangled states by the PIC even though $\mathds{P}(1)_{\text{theo}} \le 0.5$, while \textbf{FN}s are the entangled states which the PIC do not recognize. We can note that 10 and 11 states among the 2050 tested states are \textbf{FN}s and \textbf{FP}s, respectively. Since 21 states are falsely identified, the overall accuracy is $99.0(1)\%$. 
Then, we compute the Precision and Recall metrics as follows:
\begin{equation}
    \text{Precision} = \frac{\text{\textbf{TP}}}{\text{\textbf{TP}} + \text{\textbf{FP}}} = 96.3(1)\,, \quad
    \text{Recall} = \frac{\text{\textbf{TP}}}{\text{\textbf{TP}} + \text{\textbf{FN}}} = 96.6(1)\,,
\end{equation}
indicating that the circuit has the same probability of producing \textbf{FP}s \textbf{FN}s. 
\begin{figure}
    \centering
    \includegraphics[width=\linewidth]{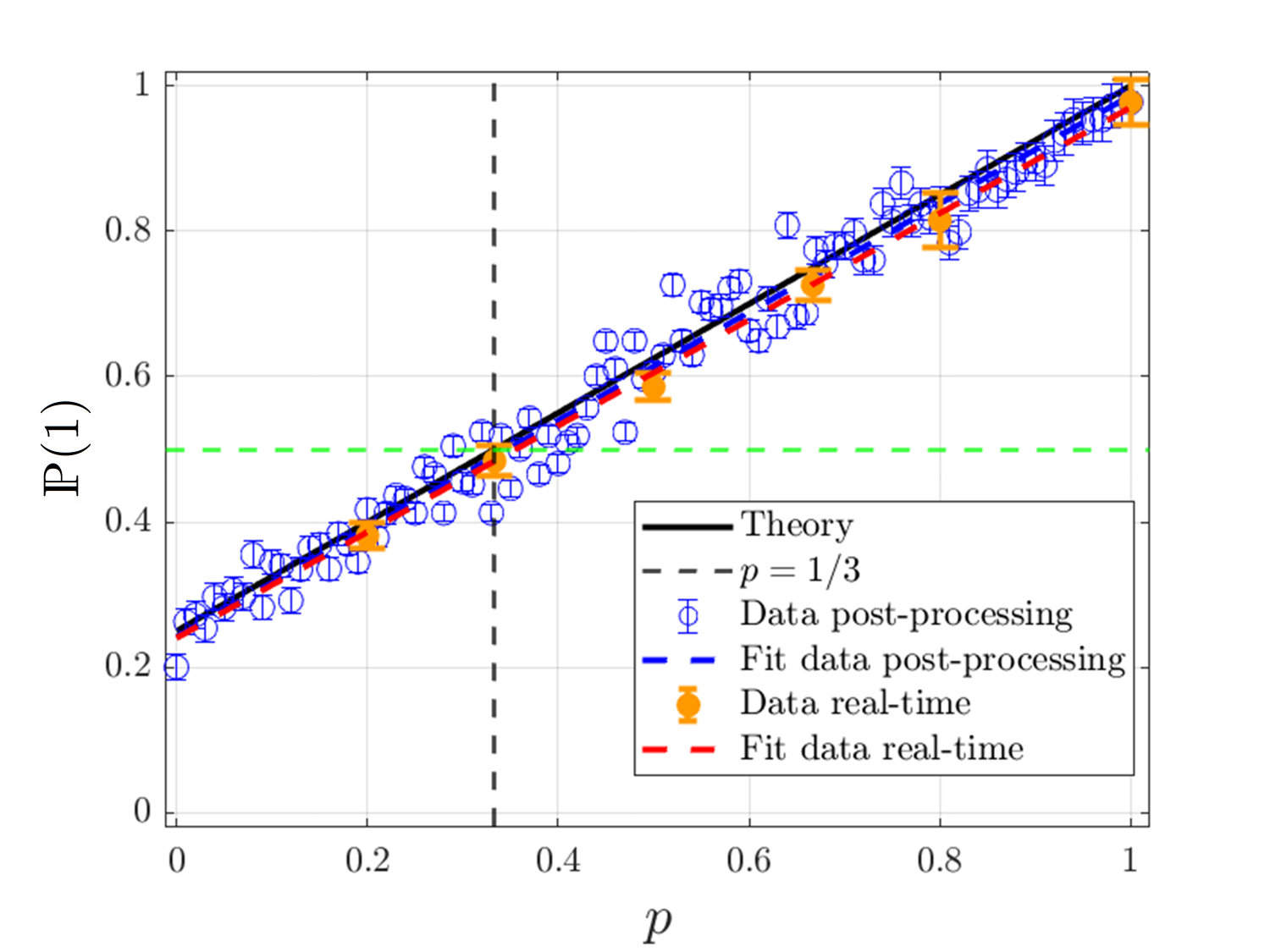}
    \caption{Results of the probability $\mathds{P}(1)$ for the family of Werner-like mixed states described by Eq. \eqref{eq:Werner-state} with $\ket{\Phi}=\ket{\Psi^-}$. Real-time experimental data with error bars are shown in orange, while data obtained through the post-postprocessing procedure described in the text are displayed in blue. The black solid line represent the theoretical prediction given by Eq. \eqref{eq:P1-werner}. A linear model $y = mx + q$ is fitted to both datasets, giving $m = 0.73(5)$ and $q = 0.24(3)$ for the real-time data, and $m = 0.742(7), \ q = 0.240(3)$ for the post-processed data.} 
    \label{fig:werner_res}
\end{figure}

Finally, we analysed the robustness of our PIC with respect to the noise which is manifesting as an uncertainty in the preparation of the state. By definition, the resulting states are mixed states, Eq.\eqref{eq:mixed_states}. We decide to focus on the analysis of the Werner-like mixed states shown in Eq. \eqref{eq:Werner-state} with $\ket{\Phi}=\ket{\Psi^-}$, Eq. \eqref{eq:bell_states}. In order to reproduce the preparation of this class of mixed states, we adopt two strategies: real-time and post-processing noise. Both ways are based on the following observation. The Werner-like mixed states contain five pure-state contributions: one coming from the state $\ket{\Phi}$, and the other four from each computational basis state\footnote{This becomes more evident if we write $\mathds{1} = \ket{00}\bra{00}+\ket{01}\bra{01}+\ket{10}\bra{10}+\ket{11}\bra{11}$.}, i.e. $\{ \ket{00},\ket{01},\ket{10},\ket{11} \}$. The probability of preparing one of these states is $p$ in the first case, and $(1-p)/4$ for the other four cases. In both strategies, the preparation of each of the five pure states is implemented by applying a specific set of electrical currents to the thermal phase shifters in the preparation stage, thereby setting the phases required to generate the corresponding state on the PIC.
The first strategy, denoted as real-time noise, consists of performing at each run of the PIC for the sampling of $\mathds{P}(1)$ a random extraction of phase setting. The possible settings are those corresponding to $\{ \ket{\Phi},\ket{00},\ket{01},\ket{10},\ket{11} \}$, with probabilities $\{p,(1-p)/4,(1-p)/4,(1-p)/4,(1-p)/4\}$. This means that the uncertainty is introduced exactly in the state preparation, and each run is preceded by a random choice between five pure states, the correct one and the computational basis states. The second strategy, denoted as post-processing noise, introduces uncertainty in the state setting after the collection of all the data for the five possible states. We choose one precise setting and sample the results individually for the five states $\{ \ket{\Phi},\ket{00},\ket{01},\ket{10},\ket{11} \}$ in order to create a data pool for each state. Then, the evaluation of $\mathds{P}(1)$ is performed by extracting with  the multinomial distribution with probabilities $\{p,(1-p)/4,(1-p)/4,(1-p)/4,(1-p)/4\}$ the outcomes from the associated five data pools. In this case, the uncertainty is introduced after the manipulation and during the data analysis. To intuitively understand the two strategies, we can imagine a Maxwell devil disturbing our experiments: in the real-time noise the devil is acting during the preparation, while in the post-processing it is inserting uncertainty to the good data by mixing them with white noise outcomes. 
Fig. \ref{fig:werner_res} reports the result for both strategies together with the linear fit, coming from the theoretical prediction shown in Eq. \eqref{eq:P1-werner}. Experimental data with both strategies are in good agreement with the theory. Moreover in the figure, the value of $p=1/3$ associated with $\mathds{P}(1)=1/2$ is highlighted, since states with $p\le 1/3$ correspond to a separable Werner state.

Our classification outcomes validate the performance of our PIC in witnessing entanglement over a large ensemble of generic pure and mixed two-qubit states. 

\section{Conclusion}
\label{sec:section_5}
\begin{table}
    \centering
    \renewcommand{\arraystretch}{1.3}
    \begin{tabular}{ p{1.8cm} p{4.8cm} p{4.8cm} }
        \hline
        \textbf{Ref.} & \textbf{Key Features} & \textbf{Limitations} \\
        \hline
        \cite{foulds2021controlled} 
        & \raggedright \textbullet\ Parallelized swap test for pure $n$-qubit states; \newline
        \textbullet\ Entanglement detection and quantification (concurrence); \newline
       \textbullet\ Robust against typical small errors.
        &  \raggedright \textbullet\ Resource demanding (requires two copies of the state and one control-qubit per state's qubits); \newline 
        \textbullet\ Theoretical; \newline
        \textbullet\ Destructive;\newline 
        \textbullet\ Pure states only. \arraybackslash \\
        \hline
         \cite{beckey2021computable}
        & \raggedright \textbullet\ Extension to multipartite states; \newline
        \textbullet\ Definition of concentratable entanglement measures.
        & \raggedright \textbullet\ Resource demanding; \newline 
        \textbullet\ Theoretical; \newline
        \textbullet\ Destructive;\newline 
        \textbullet\ Pure states only;  \arraybackslash \\
        \hline
         \cite{zhang2024controlled}
        & \raggedright \textbullet\ Extension to bipartite and multipartite mixed states; \newline
        \textbullet\ Lower bound on the concentratable entanglement of multipartite mixed states; \newline
        \textbullet\ Criterion for genuine multipartite entanglement in pure states.
        & \raggedright \textbullet\ Resource demanding; \newline
        \textbullet\ Theoretical;  \newline
        \textbullet\ Destructive. \arraybackslash \\
        \hline
        This work
        & \raggedright \textbullet\ Resource-efficient; \newline
        \textbullet\ Experimental validation; \newline
        \textbullet\ Pure and mixed states; \newline
        \textbullet\ Concurrence lower bound; \newline
        \textbullet\ Characterization of hardware noise.
        & \raggedright \textbullet\ Destructive; \newline
        \textbullet\ Concurrence lower bound; \newline
        \textbullet\ Two-qubit states. \arraybackslash \\
        \hline
    \end{tabular}
    \caption{Comparison of SWAP-based entanglement detection and quantification methods. Note that ``destructive'' refers to different mechanisms: in the first three methods, it means that the output qubits are no longer in the original input state due to swap operations (except for product states); in this work, destructiveness refers also to the photonic implementation, where information is encoded in single photons that are irreversibly detected and thus no longer available for subsequent operations.}
    \label{tab:swap_comparison}
\end{table}

This work proposes a resource-efficient method for entanglement witnessing of two-qubit systems based on the swap test. The approach is theoretically motivated and experimentally implemented on a silicon nitride PIC, specifically designed to implement the swap test algorithm. Moreover, it is applicable to both pure and mixed states, making it suitable for real-world scenarios in which quantum states are inevitably affected by noise and imperfections. The core idea is to evaluate the probability $\mathds{P}(1)$ of measuring an auxilliary qubit in the $\ket{1}$ state at the output of the circuit. Then, the condition $\mathds{P}(1) > 1/2$ is sufficient to certify entanglement. Notably, when this condition is satisfied, it is also possible to derive a lower bound on the concurrence of the input state, both for pure and mixed states. Therefore, even with limited resources, the swap test can be used to witness entanglement, offering also partial access to entanglement quantification. Furthermore, the witness is weakly optimal, as it is tangent to the set of separable states (see App. \ref{app:A}), and simple to implement, as it requires only the measurement of a single observable, i.e. $\mathds{P}(1)$. 

The main limitation of this witness is its selectivity: numerical simulations show that approximately $12.5\%$ of random entangled pure states (sampled from the unit sphere in $\mathds{C}^4$) satisfy the detection criterion. The detection capability can be enhanced to about $50 \%$ by introducing local unitary pre-processing and repeating the circuit for four appropriately selected configurations. This method results to be particularly valuable in applications where the singlet states $\ket{\Psi^-}$ serves as a resource, such as quantum teleportation \cite{bennett1993teleporting}, superdense coding \cite{bennett1992communication}, quantum key distribution \cite{ekert1991quantum}, and schemes based on decoherence-free subspaces \cite{kwiat2000experimental,lidar1998decoherence,kielpinski2002architecture}. Tab. \ref{tab:swap_comparison} compares the SWAP-based entanglement detection and quantification methods discussed in Sec.~\ref{sec:intro} with this work. In summary, our work provides a resource-efficient entanglement witness, requiring only half the qubits and gates used in previous SWAP-based schemes. The trade-off is that the protocol yields only a lower bound on the concurrence rather than a full entanglement quantification.

The proposed method is hardware-agnostic and can be implemented on various quantum platforms. Here we validated the method on a purposely designed PIC which relies solely on well-established integrated linear optical components. This results in a circuit that is simple, robust, efficient and capable of operating at room temperature with good performances. The main drawback of the photonic implementation is its reliance on single-photon quantum logic \cite{PhysRevA.57.R1477}: in the circuit, a single photon propagates through multiple waveguides, enabling the \textit{simulation} of two-qubit entangled states but not genuine physical entanglement between independent subsystems (more details are provided in App.~\ref{app:analytical_description}). Finally, the photonic design has limited scalability and the detection process is inherently destructive. 
\section*{Methods}
A schematic of the experimental setup is provided in App. \ref{app:setup}. The light source is a Ti:Sapphire laser tuned to $750$ nm. The laser output is attenuated using a variable optical attenuator (VOA), and its polarization is adjusted to be transverse electric (TE) at the chip input via a fiber polarization controller (FPC). Light is coupled into the photonic chip using a tapered lensed fiber, while the output is collected through a standard fiber array. Detection is performed using four silicon-based single-photon avalanche diodes (SPADs, Excelitas), all with equal detection efficiencies. The photon detection events are recorded by time-tagging electronics (Swabian Instruments), which are interfaced with a PC for data acquisition. A total photon flux of order of $10^6$ photons per second was used, with a time bin of $0.2 \ \mu$s set on the time tagger. The photonic circuit was fabricated via a photolithographic process through a commercial service provided by Ligentec SA foundry. The waveguides are based on a silicon nitride core, 150 nm thick and 550 nm wide, embedded in a silica cladding. These dimensions were chosen to ensure single-mode propagation at the operating wavelength of $750$ nm, for transverse electric (TE) polarization. An electrical power supply (Qontrol systems - BP8 device model) provides the currents for the phase shifters of the PIC.
\subsection*{Funding}
This project has been supported by Q@TN, the joint lab between the University of Trento, FBK - Fondazione Bruno Kessler, INFN - National Institute for
Nuclear Physics and CNR - National Research Council. A.B. acknowledges support by Horizon Widera 2023 (101160101) through ToEQPL project. 
\subsection*{Acknowledgments}
We acknowledge Matteo Sanna and Nicolò Leone for their precious contributions during the early stages of this work, with special thanks to Nicolò Leone for the design of the photonic circuit and the associated PCB.
\subsection*{Disclosures}
The authors declare no conflicts of interest.
\subsection*{Data availability} Data underlying the results presented in this paper are not publicly available at this time but may be obtained from the authors upon reasonable request.

\newpage
\appendix
\section{Derivation of the entanglement witness condition for pure states}
\label{app:A}
We show that if the condition in Eq. \eqref{eq:witnessing_condition_pure} is fulfilled, i.e. if $|\gamma - \delta| > 1$, then the Schmidt rank \cite{nielsen2010quantum} of the state $\ket{\Phi}$ must be equal to $2$, so $\ket{\Phi}$ is an entangled state. \\
Let us consider the matrix of coefficients $M$ associated to the state $\ket{\Phi}$
\begin{equation}
    M = \begin{pmatrix}
        \alpha & \delta \\
        \gamma & \beta
    \end{pmatrix}\,,
\end{equation}
then the Schmidt rank of the state $\ket{\Phi}$ is equal to the number of strictly positive eigenvalues of $M M^{\dag}$.
The eigenvalues are
\begin{equation}
    \lambda_{\pm} = \frac{1 \pm \sqrt{1-4|\alpha \beta - \gamma \delta|^2}}{2} = \frac{1 \pm \sqrt{1-C(\Phi) ^2}}{2}\,.
    \label{eigenvectors}
\end{equation}
where $C(\Phi) = 2 |\alpha \beta - \gamma \delta|$ is the concurrence of the state $\ket{\Phi}$. The eigenvalue $\lambda_+$ is always positive while $\lambda_-$ can vanish and this occurs if and only if $|\alpha \beta - \gamma \delta|^2 = 0$. In other words, the state $\ket{\Phi}$ is a product state if and only if the complex coefficients $\alpha, \beta, \gamma$ and $\delta$ are such that $|\alpha|^2 + |\beta|^2 + |\gamma|^2 + |\delta|^2 = 1$ and $|\alpha \beta - \gamma \delta|^2 = 0$. Hence, let us show that these two conditions are not compatible with Eq. \eqref{eq:witnessing_condition_pure} or, equivalently, that
\begin{equation}
    \max_{A}  \frac{1}{2}|\gamma-\delta|^2=\frac{1}{2}\,,
\end{equation}
where $A=\{(\alpha,\beta,\gamma,\delta)\in \mathds{C}^4 \ | \ |\alpha\beta-\gamma\delta|^2=0,|\alpha|^2+|\beta|^2 +|\gamma|^2 +|\delta|^2=1  \}$.\\
From the condition $|\alpha\beta-\gamma\delta|^2=0$, we trivially obtain
\begin{equation}\label{eq:ab_eq_gd}\alpha\beta=\gamma\delta\,.\end{equation}
Hence, if $\beta \neq 0$, it follows $\alpha= \frac{\gamma\delta}{\beta}$. 
On the other hand, when $\beta =0$, the condition $|\alpha\beta-\gamma\delta|^2=0$ implies that either $\gamma=0$ or $\delta =0$. In both cases, the maximum of the map $\frac{1}{2}|\gamma-\delta|^2$ is equal to $1/2$, that is,
\begin{equation}\label{bound-beta=0}
    \max_{A\cap \{\beta =0\}}  \frac{1}{2}|\gamma-\delta|^2=\frac{1}{2}\,.
\end{equation}
Thus, let us restrict ourselves to the case where $\beta \neq 0$ and $\alpha= \frac{\gamma\delta}{\beta}$. By inserting this relation into the normalization constraint we get
\begin{equation}
\be^4 +\be^2(\ga^2+\de^2-1) +\ga^2\de^2=0\,. \label{eqBeta}
\end{equation}
A necessary condition for the existence of a $|\beta|\in [0,1]$ satisfying  \eqref{eqBeta} is
\begin{equation}
    \label{eqDelta}
    \Delta=(\ga^2+\de^2-1) ^2-4\ga^2\de^2\geq0\,.
\end{equation}
The inequality is fulfilled if either $\de^2\geq\ga^2+1+2\ga$ or $\de^2\leq\ga^2+1-2\ga\,.$ The first condition is never satisfied as $\de \in [0,1]$ while the second one gives
\begin{equation}\label{cond-ga-de}
    \ga+\de\leq1\,.
\end{equation}
In other words, the set $A\cap \{\beta \neq 0\}$ is included in the set $B=\{(\alpha,\beta,\gamma,\delta):\ga+\de\leq 1\}$. Consequently,
\begin{equation} \max_{A\cap \{\beta\neq 0\}}  \frac{1}{2}|\gamma-\delta|^2\leq \max_{B}  \frac{1}{2}|\gamma-\delta|^2\,.\end{equation}
By introducing the notation 
$ \gamma =\ga e^{\text{i} \theta_\gamma}$ and $\delta =\de e^{\text{i} \theta_\delta}$, we can write
\begin{equation}\label{eq-ga-de}|\gamma-\delta|=\sqrt{\ga^2+\de^2-2\ga\de\cos \theta} \end{equation} with $\theta:=\theta_\delta-\theta_\gamma$. Since $\ga,\de$ are both positive values, we obtain
\begin{equation}\max_{\theta}\sqrt{\ga^2+\de^2-2\ga\de\cos \theta}=\sqrt{\ga^2+\de^2+2\ga\de}=\ga+\de\,.\end{equation}
Hence 
\begin{equation}\label{final-bound-pure-states}
    \max_{A\cap \{\beta\neq 0\}}  \frac{1}{2}|\gamma-\delta|^2\leq \max_{B}  \frac{1}{2}|\gamma-\delta|^2\leq \max_{B}  \frac{1}{2}(\ga+\de)^2=\frac{1}{2}\,.
\end{equation}
From Eqs. \eqref{bound-beta=0} and \eqref{final-bound-pure-states}, we conclude that
\begin{equation}\label{final-bound-pure-states-A}
    \max_{A}  \frac{1}{2}|\gamma-\delta|^2\leq \frac{1}{2}\,.
    \end{equation}
This result implies that whenever $\mathds{P}(1) > 1/2$, the state $|\Phi\rangle$ defined in Eq. \eqref{eq:generic_two_qubit_state} must be entangled. However, the condition is not necessary as an entangled state may still result in a $\mathds{P}(1) \leq 1/2$. For instance, the Bell state $\ket{\Phi^+} = \frac{1}{\sqrt{2}}(\ket{00} + \ket{11})$ yields $\mathds{P}(1) = 0$. Therefore, $\mathds{P}(1)$ can be used as an entanglement witness.

Actually, the bound in Eq. \ref{final-bound-pure-states-A} is sharp. Indeed, the point $(\alpha, \beta, \gamma, \delta) = \frac{1}{2}(1, -1, 1, -1) \in A$ gives $\frac{1}{2}|\gamma-\delta|^2 = \frac{1}{2}$. The analogous bound from below is trivial since
\begin{equation}
    \min_A \frac{1}{2}|\gamma - \delta|^2 = 0
\end{equation}
as one can easily verify by taking, e.g., $(\alpha, \beta, \gamma, \delta) = \frac{1}{2}(1,1,1,1)$.
\section{Derivation of the concurrence lower bound for pure states}
\label{app:low_bound_conc_pure}
Let us consider the case in which $\mathds{P}(1)_\Phi> 1/2$. It is convenient to introduce the positive parameter $\epsilon$ that quantifies how much the probability $\mathds{P}(1)_\Phi$ exceeds the threshold $1/2$: \begin{equation}\epsilon: =|\gamma-\delta|^2-1\,.\end{equation} We are then interested in minimizing the map $2|\alpha\beta-\gamma\delta|$ under the constraints \begin{align} &|\alpha|^2+\be^2 +\ga^2 +\de^2=1\,, \label{constraint:1}\\ & |\gamma-\delta|^2=1+\epsilon \qquad \epsilon \in (0,1]\,.\label{constraint:2} \end{align}
The optimization problem can be equivalently formulated in the minimization of the function $|\alpha\beta-\gamma\delta|^2$ under the constraints \eqref{constraint:1} and \eqref{constraint:2}. Thus, let us set \begin{equation}\label{min-concurrence-1} m(\epsilon):=\min_{{\small\begin{array}{c} |\alpha|^2+\be^2 +\ga^2 +\de^2=1 \\ |\gamma-\delta|^2=1+\epsilon \end{array}}}|\alpha\beta-\gamma\delta|^2\,.\end{equation} By introducing the notation $\alpha =\al {\rm e}^{\text{i} a}$, $\beta=\be {\rm e}^{\text{i} b}$, $\gamma =\ga {\rm e}^{\text{i} c}$, $\delta =\de {\rm e}^{\text{i} d}$, with $\al,\be,\ga,\de \in \bR^+$ and $a,b,c,d\in [0,2\pi)$, we have: \begin{equation}\label{function:2} |\alpha\beta-\gamma\delta|^2=\al ^2\be ^2+\ga^2\de^2-2\al\be\ga\de\cos(a+b-c-d)\,. \end{equation} By the explicit form of the map on the right hand side of \eqref{function:2} and the invariance of the constrains \eqref{constraint:1} and \eqref{constraint:2} under modification of the angles $a,b$, the minimization problem \eqref{min-concurrence-1} reduces to finding \begin{equation}m(\epsilon)=\min_{{\small\begin{array}{c} |\alpha|^2+\be^2 +\ga^2 +\de^2=1 \\ |\gamma-\delta|^2=1+\epsilon \end{array}}}(\al\be-\ga\de)^2\,.\end{equation} Moreover the constraint \eqref{constraint:2} can be equivalently written as \begin{equation}\ga^2+\de^2-2\ga\de \cos(c-d)=1+\epsilon\, ,\end{equation} and by setting the overall phase of the state \eqref{eq:generic_two_qubit_state} in such a way that $c=0$, it reduces to \begin{equation}\ga^2+\de^2-2\ga\de \cos(d)=1+\epsilon\,.\end{equation} Hence, \begin{equation}m(\epsilon)=\min_{{\small\begin{array}{c} |\alpha|^2+\be^2 +\ga^2 +\de^2=1 \\ \ga^2+\de^2-2\ga\de \cos(d)=1+\epsilon \end{array}}}(\al\be-\ga\de)^2\,.\end{equation} By direct computation one can verify that $m(\epsilon)=\epsilon^2/4$. Therefore, \begin{equation} \min_{\mathds{P}(1)_\Phi=\frac{1+\epsilon}{2}}2(\al\be-\ga\de)=\epsilon \end{equation} or, equivalently, \begin{equation} C(\Phi)\geq 2\mathds{P}(1)_\Phi-1 \end{equation} whenever $\mathds{P}(1)_\Phi>1/2$.
\section{Werner states}
\label{app:werner_states}
Let us consider a generic Werner-like state of the form 
\begin{equation}\label{Werner-state}
\rho=  p \ket{\Phi}\bra{\Phi} + \frac{1-p}{4} \mathds{1}\,,
\end{equation}
with $\ket{\Phi}$ a generic state vector  of the form \eqref{eq:generic_two_qubit_state}. According to the Schmidt decomposition, $\ket{\Phi}$ can be represented as
\begin{equation}
    \ket{\Phi} = \sqrt{\lambda_+} \ket{v_1 w_1} + \sqrt{\lambda_-} \ket{v_2 w_2}\,,
\end{equation}
where $\lambda_{\pm}$ are the Schmidt coefficients of the decomposition and $v_1, v_2\in \mathcal{H_A} = \mathds{C}^2$, $w_1, w_2\in \mathcal{H_B} = \mathds{C}^2$  with $\langle v_1,v_2\rangle = \langle w_1,w_2\rangle =  0$.  In particular, the Schmidt coefficients are given by
\begin{equation}
    \lambda_{\pm} = \frac{1 \pm \sqrt{1-4|\alpha \beta - \gamma \delta|^2}}{2}\,.
\end{equation}
In order to study the separability of $\rho$ by means of the PPT criterion, it is convenient to represent it  
 in the orthonormal basis of $\mathds{C}^2 \otimes \mathds{C}^2$ made by the vectors $\{\ket{v_1 w_1}, \ket{v_1 w_2}, \ket{v_2 w_1}, \ket{v_2 w_2}\}$. Exploiting the fact that the identity has the same representation in every basis, we get 
\begin{equation}
\rho =  \begin{bmatrix}
\frac{1-p}{4} + p \lambda_+ & 0 & 0 & p \sqrt{\lambda_+}  \sqrt{\lambda_-}^{*} \\
0 &  \frac{1-p}{4} & 0 & 0 \\
0 & 0 & \frac{1-p}{4}  & 0 \\
p \sqrt{\lambda_-} \sqrt{\lambda_+}^{*} & 0 & 0 & \frac{1-p}{4} + p\lambda_-
\end{bmatrix}\,.
\end{equation}
We can compute its partial transpose
\begin{equation}
\rho^{\text{PT}} =\begin{bmatrix}
\frac{1-p}{4} + p\lambda_+ & 0 & 0 & 0 \\
0 &  \frac{1-p}{4} & p \sqrt{\lambda_-} \sqrt{\lambda_+}^{*} & 0 \\
0 &  p \sqrt{\lambda_+}  \sqrt{\lambda_-}^{*} & \frac{1-p}{4}  & 0 \\
0 & 0 & 0 & \frac{1-p}{4} +  p\lambda_-
\end{bmatrix}
\end{equation}
and verify that 
\begin{equation}
    \rho \text{ is separable } \iff p \leq \frac{1}{1+4|\sqrt{\lambda_+ \lambda_-}|}\,.
\end{equation}
By computing the product
\begin{equation}
    \lambda_+ \lambda_- = \frac{1-(1-4|\alpha \beta - \gamma \delta|^2)}{4} = |\alpha \beta - \gamma \delta|^2\,,
\end{equation}
the separability condition can also be expressed as
\begin{equation}
    \rho \text{ is separable } \iff p \leq \frac{1}{1+4|\alpha \beta - \gamma \delta|} = \frac{1}{1+2C(\Phi)}\,.
\end{equation}

The probability of obtaining the state $1$ for the auxilliary qubit in the swap test circuit initially fed with the state \eqref{Werner-state} is given by:
\begin{align}
    \mathds{P}(1)& =\mathrm{Tr} \left[\mathds{1}\otimes |1\rangle\langle 1|U_{\text{SWAP}}(\rho\otimes|0\rangle\langle 0|)U_{\text{SWAP}}\right]\nonumber\\
    &=\frac{p}{2} |\gamma-\delta|^2+\frac{(1-p)}{4}\nonumber\\
    &=\frac{1}{4}+\frac{p}{2}\left(|\gamma-\delta|^2-\frac{1}{2}\right)\label{P1-werner}
\end{align}
and we can look for its minimum and maximum value in the space of separable states:
\begin{equation}m:=\min_\mathcal{S}\left[\frac{1}{4}+\frac{p}{2}\left(|\gamma-\delta|^2-\frac{1}{2}\right)\right]\,,\qquad M:=\max_\mathcal{S}\left[\frac{1}{4}+\frac{p}{2}\left(|\gamma-\delta|^2-\frac{1}{2}\right)\right]\,,\end{equation}
with $\mathcal{S}=\{(\alpha,\beta,\gamma,\delta,p)\in \mathds{C}^4\times [0,1]:|\alpha|^2+\be^2 +\ga^2 +\de^2=1, p \leq \frac{1}{1+4|\alpha \beta - \gamma \delta|}\}$ being the parameter set  associated to separable states.\\
 The minimum is trivial, i.e. $m=0$, since it is attained, e.g., for $p=1$ and $(\alpha,\beta,\gamma,\delta)=\frac{1}{2}(1,1,1,1)$.\\
The maximum of $\mathds{P}(1)$ is reached when $p\left[|\gamma-\delta|^2-\frac{1}{2}\right]$ is maximized, hence
\begin{equation}M=\frac{1}{4}+\frac{\tilde  M}{4}\,,\quad \hbox{with } \tilde M=\max_{\mathcal{S}} \left[ p\left(2|\gamma-\delta|^2-1\right) \right]\,.\end{equation}
Since $p$ is, by definition, a non-negative parameter, we get a positive value for $\tilde M$ if and only if we restrict to the subset of the parameter space $\mathcal{S}$ of those points $(\alpha,\beta,\gamma,\delta,p)$ such that $2|\gamma-\delta|^2-1>0$, 
and hence
\begin{equation}\tilde M=\max_{\mathcal{\tilde S}} \left[ p\left(2|\gamma-\delta|^2-1\right) \right]\,,\end{equation}
with 
\begin{equation}
    \tilde S=\left\{ (\alpha,\beta,\gamma,\delta,p)\in \mathds{C}^4\times [0,1]:|\alpha|^2+\be^2 +\ga^2 +\de^2=1,p \leq \frac{1}{1+4|\alpha \beta - \gamma \delta|}, |\gamma-\delta|^2\geq 1/2 \right\}\,.
\end{equation}
In particular, by the conditions  characterizing the new parameter set $\tilde S$, we have
\begin{align}
    \max_{\mathcal{\tilde S}} \left[ p\left(2|\gamma-\delta|^2-1\right) \right]&\leq \max_{\mathcal{B}\cap \{|\gamma-\delta|^2>1/2\}} \left[ \frac{1}{1+4|\alpha \beta - \gamma \delta|}\left(2|\gamma-\delta|^2-1\right) \right]\nonumber\\
    &\leq \max_{\mathcal{B}} \left[ \frac{1}{1+4|\alpha \beta - \gamma \delta|}\left(2|\gamma-\delta|^2-1\right) \right]\,,
    \end{align}
with $\mathcal{B} = \{\alpha, \beta, \gamma, \delta \ | \ |\alpha|^2 + |\beta|^2 + |\gamma|^2 + |\delta|^2 = 1  \} $. By introducing the notation $\alpha = |\alpha| {\rm e}^{\text{i} a}$, $\beta = |\beta| {\rm e}^{\text{i} b}$, $\gamma = |\gamma| {\rm e}^{\text{i} c}$, $\delta = |\delta| {\rm e}^{\text{i} d}$, we can write
\begin{multline}
     \frac{1}{1+4|\alpha \beta - \gamma \delta|}\left(2|\gamma-\delta|^2-1 \right)=\\=\frac{2\ga^2+2\de^2-4\ga\de \cos(c-d)-1}{(1+4\sqrt{\al^2\be^2+\ga^2\de^2-2\al\be\ga\de\cos(a+b-c-d)})}\,.
\end{multline}
Thus, by choosing the values of the angles $a,b,c,d$ that maximize the numerator and minimize the denominator, 
we get
\begin{equation} \max_{\mathcal{B}} \left[ \frac{1}{1+4|\alpha \beta - \gamma \delta|}\left(2|\gamma-\delta|^2-1\right)\right]=\max_{\mathcal{B}}\frac{2(\ga+\de)^2-1}{(1+4\left|\al\be-\ga\de\right|)}\,.\end{equation}
By symmetry and numerical analysis, we expect
\begin{equation}
     \max_{\mathcal{B}} \left[  \frac{1}{1+4|\al \be - \ga \de|} \left(2|\ga+\de|^2-1\right) \right] = 1\,,
    \label{eq:max_sep_generic_case}
\end{equation}
and this maximum is attained for $\ga=\de=\sqrt 2/2$ and $\al=\be=0$.
In order to prove \eqref{eq:max_sep_generic_case}, it is sufficient to note that it is equivalent to showing that:
\begin{equation}\label{eq-maximum-real-case}
    \max_{D}\left[x^2+y^2+2xy-2|zw-xy|\right]=1,
\end{equation}
with $D=\{(x,y,z,w)\in \mathds{R}^4:x^2+y^2+z^2+w^2=1, x\geq 0,y\geq 0,z\geq 0,w\geq 0\}$. Denoted by $\bar M $ the left hand side of Eq.\eqref{eq-maximum-real-case}, one has that $\bar M=\max\{M_1,M_2\}$ with
\begin{align}
    &M_1:= \max_{D_1}\left[x^2+y^2+2zw\right]\qquad D_1=D\cap \{zw\leq xy\}\\
    &M_2:= \max_{D_2}\left[x^2+y^2+4xy-2 zw\right]\qquad D_2=D\cap \{zw> xy\}
\end{align}
Both values can be easily computed by applying Lagrange multipliers, yielding the result $\bar M=1$. 

\section{The quantum min-entropy and its certification via the $\mathds{P}>1/2$ inequality.}
\label{app:qu_randomness}
Let us consider the case in which the input state  of the swap test is $\ket{\Phi} \otimes \ket{0}$, where $\ket{\Phi}$ is the generic two-qubit pure state defined in Eq. \eqref{eq:generic_two_qubit_state}. According to Theorem \ref{theo:ent_witness_pure}, if $\mathds{P}(1)>1/2$ then $\ket{\Phi}$ must be entangled and its Schmidt decomposition can be written in the form
\begin{equation}\label{Sc-dec}
    \ket{\Phi}=\cos\theta \ket{v_1w_1}+\sin\theta \ket{v_2w_2}\,,
\end{equation}
with $\theta \in (0,\pi/4]$, $\{v_1,v_2\}$ and $\{w_1,w_2\}$ orthonormal basis of $\mathds{C}^2$.\\
Any local observable has the form $O^{\ba}:= \ba \cdot \sigma \otimes \mathds{1}$ or $O^{ \bb}=\mathds{1}\otimes  \bb \cdot \sigma$, with $ \ba, \bb$ unit vectors of $\mathbb{R}^3$ and $v \cdot \sigma:=v_x\sigma_x+v_y\sigma_y+v_z\sigma_z$. Clearly, the measurements of $O^{ \ba}$ and $O^{ \bb}$ each have two possible outcomes, $\pm 1$, with corresponding probabilities
\begin{align}\label{detection-probabilities-local-a}
\mathds{P}(x| \Phi,  \ba)&=\Tr\left[|\Phi \rangle \langle \Phi | \left(\frac{1}{2}(\mathds{1}+x  \ba \cdot \sigma)\otimes \mathds{1}\right)\right]\,,\qquad x=\pm 1\,,\\
\mathds{P}(y|\Phi,  \bb)&=\Tr\left[|\Phi \rangle \langle \Phi | \left(\mathds{1}\otimes \frac{1}{2}(\mathds{1}+y \bb \cdot \sigma)\right)\right]\,,\qquad y=\pm 1\,. \label{detection-probabilities-local-b}
\end{align}
In the following we are interested in studying the so-called {\em guessing probability}, i.e. the probability of the most likely outcome, defined as
\begin{align}
G(\Phi,  \ba)&:=\max_{x=\pm 1}\mathds{P}(x|\Phi,  \ba)\,,\\
G(\Phi,  \bb)&:=\max_{y=\pm 1}\mathds{P}(y|\Phi,  \bb)\,,
\end{align}
and, in a device-independent framework, compute its maximum values over all choices of local observables
\begin{equation}\label{guessing-probability-DI-pure}
G(\Phi):=\max\{\max _{ \ba}G(\Phi,  \ba),\max _{\bb}G(\Phi, \bb) \}\,.
\end{equation}
This figure of merit provides an overall measure of the predictability of the outcomes of the quantum measurement, independently of the particular choice of the local observable that is measured \cite{herrero2017quantum}.\\
The detection probabilities \eqref{detection-probabilities-local-a} and \eqref{detection-probabilities-local-b} can be easily computed in the case where $v_1,v_2$ and $w_1,w_2$  coincide with the computational basis $|0\rangle, |1\rangle$. Indeed, in this case, 
\begin{equation}\label{expectation-local-observables}
\mathrm{Tr}\left[|\Phi\rangle\langle \Phi | O^{ \ba}\right] =  a _z \cos 2\theta\,, \qquad \mathrm{Tr}\left[|\Phi\rangle\langle \Phi|  O^{ \bb}\right] =  b _z \cos 2\theta\,, 
 \end{equation}
More generally, by introducing the local unitary $U\otimes V$ such that 
\begin{equation}|v_1\rangle=U|0\rangle\,,\, |v_2\rangle=U|1\rangle\,, \qquad |w_1\rangle=V|0\rangle\,,\, |w_2\rangle=V|1\rangle\,,\end{equation}
and by denoting by $\tilde \ba ,\tilde\bb$ the unit vectors in $\mathbb{R}^3$ such that
\begin{equation}\tilde a\cdot \sigma =U^+ a\cdot \sigma U\,, \qquad \tilde b\cdot \sigma =V^+ b\cdot \sigma V\,,\end{equation}
Eq. \eqref{expectation-local-observables} still holds with $a_z$ and $b_z$ replaced by $\tilde a_z$ and $\tilde b_z$ respectively. Hence, the guessing probability \eqref{guessing-probability-DI-pure} is equal to
\begin{equation}\label{guessing-probability-DI-pure-entangledstates}
G(\Phi)=\frac{1+\cos(2\theta)}{2}=\cos^2\theta\,.
\end{equation}\\
According to  Eq. \eqref{eigenvectors}, for a general state of the form \eqref{eq:generic_two_qubit_state}, the following identity holds:
\begin{align}
    \cos^2\theta&=\frac{1+\sqrt{1-4|\alpha\beta-\gamma\delta|^2}}{2}
    =\frac{1+\sqrt{1-C(\Phi)^2}}{2}\,.
\end{align} 
Theorem \eqref{theo:conc_bound} states that the value $\mathds{P}(1)_\Phi$ of the detection probability of the auxiliary qubit state $\ket{1}$ in the swap test circuit provides a lower bound for the concurrence of the state $|\Phi\rangle$. This yields an upper bound for the quantum guessing probability of the following form
\begin{equation} \label{eq:bound_G_pure}
    G(\Phi)\leq \frac{1+\sqrt{1-f^2(\mathds{P}(1)_\Phi)}}{2}\,,
\end{equation}
where $f:[0,1]\to \mathbb{R}$ is in general given by \eqref{eq:function_bound_non_ideal}.

This result can be generalized to arbitrary mixed states. In the case of the measurement of a local observable of the form $O^{\ba}:= \ba \cdot \sigma \otimes \mathds{1}$ or $O^{ \bb}=\mathds{1} \otimes  \bb \cdot \sigma$ over a bipartite mixed state $\rho$ of two qubits, the quantum guessing probability is defined as the average 
   \begin{align}
       G(\rho, \ba)&:=\sup_{\{(\mu_\lambda,\psi_\lambda)\}}\int_\Lambda   G(\psi_\lambda, \ba) d\mu(\lambda)\,,\\
        G(\rho, \bb)&:=\sup_{\{(\mu_\lambda,\psi_\lambda)\}}\int_\Lambda   G(\psi_\lambda, \bb) d\mu(\lambda)\,,
   \end{align}
  where the supremum is taken over all possible choices of pairs  $\{(\mu(\lambda), \psi_\lambda)_{\lambda \in \Lambda}\}$ such that
  \begin{equation}\label{mixture}\rho=\int_\Lambda  |\psi _\lambda\rangle \langle  \psi _\lambda | d\mu(\lambda)\:,\end{equation} in {\em weak sense}, where $\mu$ is any probability measure over the space $\Lambda$ and the vectors $\psi_\lambda$ are not mutually orthogonal in general.  In a nutshell  $G(\rho, \ba)$ is interpreted as the average  guessing probability given the knowledge on which particular pure state in the mixture \eqref{mixture} has been prepared. In this case the additional (classical) stocasticity present in the mixed state $\rho$ is interpreted as ``side information'', which is potentially accessible to an adversary and has to be suitably taken into account in the construction of a figure of merit able to quantify the amount of secure (i.e., unpredictable) quantum randomness.\\
  According to Eq. \eqref{eq:bound_G_pure}, if $\psi _\lambda$ is a pure state yielding a probability $\mathds{P}(1)_{\psi_\lambda}$, then for any factorized observable $\ba\cdot \sigma \otimes \mathds{1}$ (or  $\mathds{1}\otimes \bb \cdot \sigma $) the guessing probability is bounded by
  \begin{equation}
      G(\psi_\lambda, \ba) \leq g(\mathds{P}(1)_{\psi_\lambda})\,, \qquad  G(\psi_\lambda, \bb) \leq g(\mathds{P}(1)_{\psi_\lambda})\,,
  \end{equation}
  where $g:[0,1]\to [1/2,1]$ is the map defined by
  \begin{equation}
  g(x):=\frac{1+\sqrt{1-f^2(x)}}{2}\,.
  \end{equation}
  In particular $g(x)=1$ for $x\leq (1+c)/2$, while $g(x)=\frac{1+\sqrt{1-(2c-1)^2}}{2}$ for $x>(1+c)/2$ and it is a concave function.

 In the general case of a mixed two-qubit state $\rho$ such that $\mathds{P}(1)_{\rho}=b$ we have that for any choice of a local observable the corresponding guessing probability is bounded by
 \begin{align}
     G(\rho, \ba)&=\sup_{\{(\mu_\lambda,\psi_\lambda)\}}\int_\Lambda   G(\psi_\lambda, \ba)d\mu(\lambda)\nonumber\\
     &\leq \sup_{\{(\mu_\lambda,\psi_\lambda)\}}\int_\Lambda g(\mathds{P}(1)_{\psi_\lambda})d\mu(\lambda)\nonumber\\
     &\leq \sup_{\{(\mu_\lambda,\psi_\lambda)\}}g\left(\int_\Lambda \mathds{P}(1)_{\psi_\lambda} d\mu(\lambda)\right)=g(\mathds{P}(1)_{\rho})
 \end{align}
 and this provides a (semi)-device independent certification of quantum randomness.
 
\section{Analytical description of the PIC functionality}
\label{app:analytical_description}
\subsection{Preparation stage}
\label{app:preparation_stage}
There are, in principle, different paradigms that can be exploited for quantum photonic computation \cite{romero2024photonic,knill2001scheme, nielsen2004optical, browne2005resource,kok2007linear}. Here, the main idea is to use single photons and integrated linear optical components only. To encode states in the PIC, path-encoded single photons are used, where the state is determined by the waveguide through which a single photon travels. For instance, if we have $2^n$ waveguides and a single photon, the system behaves as a qudit \cite{wang2020qudits} with dimension $2^n$. We can identify the state $\ket{i}$, with $i \in 0, ... , 2n-1$, with the photon being in the $i$-th waveguide. In the simple case of the swap test algorithm, the input qubits are three so $2^3$ optical paths are necessary. The encoding is achieved by converting the waveguide numbers to binary 
\begin{align}
    \ket{0} \mapsto \ket{000} &\equiv \ket{0} \otimes \ket{0} \otimes \ket{0}\,, \quad \ket{1} \mapsto \ket{001} \equiv \ket{0} \otimes \ket{0} \otimes \ket{1}\,, \\
    \ket{2} \mapsto \ket{010} &\equiv \ket{0} \otimes \ket{1} \otimes \ket{0}\,, \quad \ket{3} \mapsto \ket{011} \equiv \ket{0} \otimes \ket{1} \otimes \ket{1}\,,\\
    \ket{4} \mapsto \ket{100} &\equiv \ket{1} \otimes \ket{0} \otimes \ket{0}\,, \quad \ket{5} \mapsto \ket{101} \equiv \ket{1} \otimes \ket{0} \otimes \ket{1}\,,\\
    \ket{6} \mapsto \ket{110} &\equiv \ket{1} \otimes \ket{1} \otimes \ket{0}\,, \quad \ket{7} \mapsto \ket{111} \equiv \ket{1} \otimes \ket{1} \otimes \ket{1} \,,
\end{align}
where the states on the left-hand side of the arrow represents the qudit states, while the states on the right-hand side of the arrows represent the two qubits whose scalar product we are interested in, and the auxilliary qubit.\\
The manipulation of the quantum states is achieved by three integrated MZIs and six PSs, divided into two layers. The first MZI and the first layer of PSs (yellow box in Fig. \ref{fig:SWAP_test_photonic_gate_repr}(b)), followed by the second and third MZI and the second layer of PSs (green box in Fig. \ref{fig:SWAP_test_photonic_gate_repr}(b)) allows to establish the state of the first two qubits. \\
To gain a clearer understanding, recall that in the \textit{dual-rail encoding} scheme a single qubit is represented by a pair of waveguides, where the states $\ket{0}$ and $\ket{1}$ correspond to a single photon propagating in the upper and lower path, respectively. Within this framework, any generic unitary operation $U$ acting on the qubit’s two-dimensional Hilbert space can be implemented using a MZI and two additional PSs. More in detail if we have two waveguides and we start with one photon in the state $\ket{0}$, by applying the transformation
\begin{equation}
\label{eq:Urot}
    U_{\text{rot}}(\boldsymbol{\theta}, \boldsymbol{\phi}) \equiv U_{\text{PS}}(\boldsymbol{\phi}) \cdot U_{\text{MZI}}(\boldsymbol{\theta})\,,
\end{equation}
where
\begin{align}
    U_{\text{PS}}(\boldsymbol{\phi}) &= \begin{pmatrix}
        {\rm e}^{\text{i} \phi(1)} & 0 \\
        0 & {\rm e}^{\text{i} \phi(2)}
    \end{pmatrix}\,, \\
    U_{\text{MZI}}(\boldsymbol{\theta}) &= \text{i} {\rm e}^{\text{i} \left(\frac{\theta(1)+\theta(2)}{2}\right)}\begin{pmatrix}
        \sin{\left(\frac{\theta(1)-\theta(2)}{2}\right)} & \cos{\left(\frac{\theta(1)-\theta(2)}{2}\right)} \\
        \cos{\left(\frac{\theta(1)-\theta(2)}{2}\right)} & -\sin{\left(\frac{\theta(1)-\theta(2)}{2}\right)}
    \end{pmatrix}\,,
\end{align}
it is possible to reach any point on the Bloch sphere associated to the single qubit. In the case of multiple waveguides and a qudit structure, these transformations can be generalized and the matrix representation of a MZI followed by PSs that operates on the $k$-th and $(k+1)$-th waveguides reads 
\begin{equation}
    U_{\text{rot}}^{(k)}(\boldsymbol{\theta}, \boldsymbol{\phi}) \equiv \begin{pmatrix}
        1 & 0 & \cdots & \cdots & \cdots & 0 \\
        0 & \ddots & \ & \ & \ & \vdots \\
        \vdots & \ & (U_{\text{rot}})_{11} & (U_{\text{rot}})_{12} & \ & \vdots \\
        \vdots & \ & (U_{\text{rot}})_{21} & (U_{\text{rot}})_{22} & \ & \vdots \\
         \vdots & \ & \ & \ & \ddots & 0 \\
         0 & \cdots & \cdots & \cdots & 0 & 1 \\
    \end{pmatrix}\,.
\end{equation}
In this expression, the entries of the matrix $U_{\text{rot}}$ are determined by Eq. \eqref{eq:Urot}, and the mapping from Dirac notation to vector notation follows the conventional approach. \\
The action of the triangular arrangement of MZIs and PSs on the input photon state $\ket{01}$ (see Fig. \ref{fig:SWAP_test_photonic_gate_repr}(b)) allows the preparation of any two-qubit state, including entangled states. Specifically, let \( \boldsymbol{\theta}_1, \boldsymbol{\phi}_1 \) denote the phases of \(\text{MZI}_1\) and \(\text{PS}_1\), respectively. Similarly, let \( \boldsymbol{\theta}_{21}, \boldsymbol{\phi}_{21} \) denote the phases of \(\text{MZI}_2\) and the first pair of phase shifters in \(\text{PS}_2\), and let \( \boldsymbol{\theta}_{22}, \boldsymbol{\phi}_{22} \) denote the phases of \(\text{MZI}_3\) and the second pair of phase shifters in \(\text{PS}_2\). Then, we have
\begin{equation}
    \begin{split}
        &U_{\text{rot}}^{(3)}(\btheta_{22}, \bphi_{22}) \cdot  U_{\text{rot}}^{(1)}(\btheta_{21}, \bphi_{21}) \cdot  U_{\text{rot}}^{(2)}(\btheta_{1}, \bphi_{1}) \cdot \begin{pmatrix}
            0 \\ 1 \\ 0 \\ 0
        \end{pmatrix} = \\
         &= -{\rm e}^{-i(\theta_1(1) + \theta_1(2))} \begin{pmatrix}
            {\rm e}^{\text{i} (\phi_1(1)-(\theta_{21}(1)+\theta_{21}(2)) + \phi_{21}(1))} \sin \Delta \theta_1 \cos \Delta \theta_{21} \\
            -{\rm e}^{\text{i} (\phi_1(1)-(\theta_{21}(1)+\theta_{21}(2)) + \phi_{21}(2))} \sin \Delta \theta_1 \sin \Delta \theta_{21} \\
            {\rm e}^{\text{i} (\phi_1(2)-(\theta_{22}(1)+\theta_{22}(2)) + \phi_{22}(1))} \cos \Delta \theta_1 \sin \Delta \theta_{22} \\
             {\rm e}^{\text{i} (\phi_1(2)-(\theta_{22}(1)+\theta_{22}(2)) + \phi_{22}(2))} \cos \Delta \theta_1 \cos \Delta \theta_{22} \\
         \end{pmatrix}\,,
    \end{split}
\end{equation}
that can also be written in Dirac notation as
\begin{equation}
    \begin{split}
       &U_{\text{rot}}^{(3)}(\btheta_{22}, \bphi_{22}) \cdot  U_{\text{rot}}^{(1)}(\btheta_{21}, \bphi_{21}) \cdot  U_{\text{rot}}^{(2)}(\btheta_{1}, \bphi_{1}) \cdot \ket{01} = \\
       &= -{\rm e}^{-\text{i}(\theta_1(1) + \theta_1(2))}   \\ & \ \ \left[{\rm e}^{\text{i} (\phi_1(1)-(\theta_{21}(1)+\theta_{21}(2)) + \phi_{21}(1))} \sin \Delta \theta_1 \left(\cos \Delta \theta_{21} \ket{00} - {\rm e}^{-\text{i}\Delta \phi_{21}}\sin \Delta \theta_{21} \ket{01} \right) \right. \\
       & \ \ \left. + {\rm e}^{\text{i} (\phi_1(2)-(\theta_{22}(1)+\theta_{22}(2)) + \phi_{22}(1))} \cos \Delta \theta_1  \left(\sin \Delta \theta_{22} \ket{10} + {\rm e}^{-\text{i}\Delta \phi_{22}}\cos \Delta \theta_{22} \ket{11} \right) \right]\,,
    \end{split}
    \label{eq:generic_prepared_state_full}
\end{equation}
where $\Delta \phi_j \equiv \phi_j(1)-\phi_j(2)$. These expressions explicitly show that any two-qubit state of the form $\alpha \ket{00} + \beta \ket{11}+  \gamma \ket{10} + \delta \ket{01}$ with $|\alpha|^2 + |\beta|^2 + |\gamma|^2 + |\delta|^2 = 1$ can be set. It is worth noting that the condition for separability requires $\Delta \theta_2 \equiv \Delta \theta_{21}-\frac{\pi}{2} = \Delta \theta_{22}$ and $\bphi_2 \equiv \bphi_{21} = \bphi_{22}$. Under this condition, the two-qubit state, up to a global phase, takes the form
\begin{equation}
        ({\rm e}^{-\text{i}(\theta_{21}^{\text{(sum)}}-\theta_{22}^{\text{(sum)}})}\sin \Delta \theta_1 \ket{0} + {\rm e}^{-\text{i}\Delta \phi_1}\cos \Delta \theta_1 \ket{1}) \otimes (\sin \Delta \theta_2 \ket{0} + {\rm e}^{-\text{i}\Delta \phi_2}\cos \Delta \theta_2 \ket{1}) 
\end{equation}
where $\theta_{k}^{\text{(sum)}} = \theta_{k}(1) + \theta_{k}(2)$. In practice, separable states can be generated by choosing either $\theta_{22}(1) = \theta_{21}(1)-\frac{\pi}{2}$ and $\theta_{22}(2) = \theta_{21}(2)$, or $\theta_{22}(1) = \theta_{21}(1)$ and $\theta_{22}(2) = \theta_{21}(2)-\frac{\pi}{2}$. In both cases, by introducing $\Delta \tilde{\phi_1} = \Delta \phi_1+\frac{\pi}{2}$ and dropping global phases, the resulting state simplifies to
\begin{equation}
    \begin{split}
        (\sin \Delta \theta_1 \ket{0} +{\rm e}^{-i\Delta \tilde{\phi_1}}\cos \Delta \theta_1 \ket{1}) \otimes (\sin \Delta \theta_2 \ket{0} + {\rm e}^{-i\Delta \phi_2}\cos \Delta \theta_2 \ket{1})\,,
    \end{split}
    \label{eq:generic_sep_state}
\end{equation}
which corresponds to the general form of a separable two-qubit state.
In this configuration, the first MZI (along with the first layer of PSs) prepares the state $\ket{\psi}$ of the first qubit, while the second and third MZIs, followed by the second layer of PSs, define the state $\ket{\xi}$ of the second qubit. \\
In the final section of the preparation stage, highlighted within the blue box in Fig. \ref{fig:SWAP_test_photonic_gate_repr}(b), four additional waveguides are introduced to realize the desired $2^3$-qudit. Connecting the outputs of $\text{PS}_2$ to the even-numbered waveguides passively initialises the auxiliary qubit to the state $\ket{0}$. Indeed, according to the encoding introduced at the beginning of the section, no photons will ever be present in those waveguides corresponding to the third qubit in state $\ket{1}$. Therefore, the total state at the end of the preparation stage reads $\ket{\Psi} = \ket{\Phi} \otimes \ket{0}$ where $\ket{\Phi}$ is the generic two-qubit state defined by Eq. \eqref{eq:generic_prepared_state_full}. When the condition for separability is met this total state becomes $\ket{\Psi} = \ket{\psi} \otimes \ket{\xi} \otimes \ket{0}$ with $\ket{\psi} \otimes \ket{\xi}$ defined by Eq. \eqref{eq:generic_sep_state}. 
\subsection{Swap test stage}
The second stage of the circuit is the SWAP stage, where the test is implemented. The first layer consists of a series of MMIs, each connecting an even waveguide to an odd waveguide. The action of these MMIs is represented by the block-diagonal matrix
\begin{equation}
    \frac{1}{\sqrt{2}} \begin{pmatrix}
        1 & \text{i} & 0 & 0 & 0 & 0 & 0 & 0 \\
        \text{i} & 1 & 0 & 0 & 0 & 0 & 0 & 0 \\
        0 & 0 & 1 & \text{i} & 0 & 0 & 0 & 0 \\
        0 & 0 & \text{i} & 1 & 0 & 0 & 0 & 0 \\
        0 & 0 & 0 & 0 & 1 & \text{i} & 0 & 0 \\
        0 & 0 & 0 & 0 & \text{i} & 1 & 0 & 0 \\
        0 & 0 & 0 & 0 & 0 & 0 & 1 & \text{i} \\
        0 & 0 & 0 & 0 & 0 & 0 & \text{i} & 1 \\
    \end{pmatrix}\,,
\end{equation}
where the standard mapping between Dirac and vector notation has been assumed. This is equivalent to applying the transformation $\mathds{1} \otimes U_{\text{MMI}}$ to the total state $\ket{\Psi}$. Thus, we are acting only on the auxiliary qubit using $U_{\text{MMI}}$ instead of H-gates. \\
The core of this stage is constituted by a network of waveguide crossings, which implement the CSWAP gate. As illustrated by the purple box Fig. \ref{fig:SWAP_test_photonic_gate_repr}(b), the network consists of three CRs realizing the swapping of $\ket{011}$ and $\ket{101}$, modulo overall phases. The matrix that describes their action is as follows
\begin{equation}
    U_{\text{CSWAP}} = \begin{pmatrix}
        1 & 0 & 0 & 0 & 0 & 0 & 0 & 0 \\
        0 & 1 & 0 & 0 & 0 & 0 & 0 & 0 \\
        0 & 0 & 1 & 0 & 0 & 0 & 0 & 0 \\
        0 & 0 & 0 & 0 & 0 & 1 & 0 & 0 \\
        0 & 0 & 0 & 0 & 1 & 0 & 0 & 0 \\
        0 & 0 & 0 & 1 & 0 & 0 & 0 & 0 \\
        0 & 0 & 0 & 0 & 0 & 0 & 1 & 0 \\
        0 & 0 & 0 & 0 & 0 & 0 & 0 & 1 \\
    \end{pmatrix}\,.
\end{equation}
The subsequent layer of phase shifters (PS$_3$), introduced on all eight waveguides, is necessary to correct spurious phases between the optical paths. Indeed spurious phases can result in significant estimation errors, that depend on the pair of states, and their presence is practically inevitable due to the finite resolution of the fabrication process. The matrix representing the layer $\text{PS}_3$ is diagonal with different phases as entries
\begin{equation}
\hspace*{-1cm}
    U_{\text{PS}_3}(\btheta_s) = \begin{pmatrix}
        {\rm e}^{\text{i} \theta_s(1)} & 0 & 0 & 0 & 0 & 0 & 0 & 0 \\
        0 & {\rm e}^{\text{i} \theta_s(2)} & 0 & 0 & 0 & 0 & 0 & 0 \\
        0 & 0 & {\rm e}^{\text{i} \theta_s(3)} & 0 & 0 & 0 & 0 & 0 \\
        0 & 0 & 0 & {\rm e}^{\text{i} \theta_s(4)} & 0 & 0 & 0 & 0 \\
        0 & 0 & 0 & 0 & {\rm e}^{\text{i} \theta_s(5)} & 0 & 0 & 0 \\
        0 & 0 & 0 & 0 & 0 & {\rm e}^{\text{i} \theta_s(6)} & 0 & 0 \\
        0 & 0 & 0 & 0 & 0 & 0 & {\rm e}^{\text{i} \theta_s(7)} & 0 \\
        0 & 0 & 0 & 0 & 0 & 0 & 0 & {\rm e}^{\text{i} \theta_s(8)} \\
    \end{pmatrix}\,.
\end{equation}
An additional layer of MMIs, implementing the transformation $\mathds{1} \otimes U_{\text{MMI}}$, is also present in the final section of the stage, right before the detection.\\
Overall, the unitary operator describing the entire swap stage is
\begin{equation}
    U_{\text{swaptest}}(\btheta_s) = (\mathds{1} \otimes U_{\text{MMI}}) \cdot U_{\text{PS}_3}(\btheta_s) \cdot U_{\text{CSWAP}} \cdot (\mathds{1} \otimes U_{\text{MMI}})\,.
\end{equation}
This matrix can be factorized in three mutually commuting contributions:
\begin{equation}
\hspace*{-1.5cm}     U_{\text{MZI}}^{(1)}\left(\frac{\theta_s(1)}{2},\frac{\theta_s(2)}{2}\right) \cdot U_{\text{MZI}}^{(7)}\left(\frac{\theta_s(7)}{2},\frac{\theta_s(8)}{2}\right) \cdot U_{\text{SWAPcore}}(\theta_s(3),\theta_s(4), \theta_s(5),\theta_s(6))\,,
\end{equation}
where $ U_{\text{MZI}}^{(1)}$ represents the MZI realized between the first and second waveguide, $U_{\text{MZI}}^{(7)}$ the MZI between the seventh and the eighth waveguide while $U_{\text{SWAPcore}}(\theta_s(3),\theta_s(4), \theta_s(5),\theta_s(6))$ is defined as
\begin{equation}
\frac{1}{2}
    \begin{pmatrix}
        2 & 0 & 0 & 0 & 0 & 0 & 0 & 0 \\
        0 & 2 & 0 & 0 & 0 & 0 & 0 & 0 \\
        0 & 0 & {\rm e}^{\text{i} \theta_s(3)} & \text{i}{\rm e}^{\text{i} \theta_s(3)} &-{\rm e}^{\text{i} \theta_s(4)} & \text{i}{\rm e}^{\text{i} \theta_s(4)} & 0 & 0 \\
        0 & 0 & \text{i}{\rm e}^{\text{i} \theta_s(3)} & -{\rm e}^{\text{i} \theta_s(3)} & \text{i}{\rm e}^{\text{i} \theta_s(4)} & {\rm e}^{\text{i} \theta_s(4)} & 0 & 0 \\
        0 & 0 & -{\rm e}^{\text{i} \theta_s(6)} & \text{i}{\rm e}^{\text{i} \theta_s(6)} & {\rm e}^{\text{i} \theta_s(5)} & \text{i}{\rm e}^{\text{i} \theta_s(5)} & 0 & 0 \\
        0 & 0 & \text{i}{\rm e}^{\text{i} \theta_s(6)} & {\rm e}^{\text{i} \theta_s(6)} & \text{i}{\rm e}^{\text{i} \theta_s(5)} & -{\rm e}^{\text{i} \theta_s(5)} & 0 & 0 \\
        0 & 0 & 0 & 0 & 0 & 0 & 2 & 0 \\
        0 & 0 & 0 & 0 & 0 & 0 & 0 & 2 \\  
    \end{pmatrix}\,.
\end{equation}
Setting all the phases $\theta_s(k) = 0$ and applying $U_{\text{swaptest}}$ to the generic separable state $\ket{\Psi} = \ket{\psi} \otimes \ket{\xi} \otimes \ket{0}$ with $\ket{\psi} \otimes \ket{\xi}$ defined by Eq. \eqref{eq:generic_sep_state}, it is possible to obtain the total state at the output of the circuit \cite{baldazzi2024linear}. According to the path-encoding, the probability $\mathds{P}(0)$ of measuring the ancilla in the state $\ket{0}$ is given by the sum of the squared moduli of the even-indexed entries of the resulting state vector, while $\mathds{P}(1)$, corresponding to the ancilla found in state $\ket{1}$ is obtained from the odd-indexed entries. In particular, the expressions for $\{\mathds{P}(x)\}_{x = 0,1}$ are
\begin{align}
    \mathds{P}(0) &= \frac{1}{2}\left[1-\cos^2\left(\Delta \theta_1 - \Delta \theta_2 \right) \cos^2 \left( \frac{\Delta \tilde{\phi_1}- \Delta \phi_2}{2}\right) - \cos^2\left(\Delta \theta_1 + \Delta \theta_2 \right) \sin^2 \left( \frac{\Delta \tilde{\phi_1} - \Delta \phi_2}{2}\right) \right]\,, \\
    \mathds{P}(1) &= \frac{1}{2}\left[1+\cos^2\left(\Delta \theta_1 - \Delta \theta_2 \right) \cos^2 \left( \frac{\Delta \tilde{\phi_1} - \Delta \phi_2}{2}\right) + \cos^2\left(\Delta \theta_1 + \Delta \theta_2 \right) \sin^2 \left( \frac{\Delta \tilde{\phi_1} - \Delta \phi_2}{2}\right) \right]\,.
\end{align}
The theoretical result for the scalar product of $\ket{\psi}$ and $\ket{\xi}$ defined by Eq. \eqref{eq:generic_sep_state} can be expressed as
\begin{equation}
\label{eq:modulus_square_theoretical}
    |\langle \psi | \xi \rangle|^2 = \cos^2\left(\Delta \theta_1 - \Delta \theta_2 \right) \cos^2 \left( \frac{\Delta \tilde{\phi_1} - \Delta \phi_2}{2}\right) + \cos^2\left(\Delta \theta_1 + \Delta \theta_2 \right) \sin^2 \left( \frac{\Delta \tilde{\phi_1} - \Delta \phi_2}{2}\right)\,.
\end{equation}
Thus, the probabilities $\{\mathds{P}(x)\}_{x = 0,1}$ can also be written as
\begin{equation}
    \mathds{P}(x) = \frac{1}{2}\big[1 - (-1)^x|\langle \psi| \xi \rangle|^2 \big]\,,
\end{equation}
which corresponds to the result of the standard swap test algorithm reported in Eq. \eqref{eq:swaptest_res}, except for the exchange of $\mathds{P}(0)$ and $\mathds{P}(1)$. Hence, the photonic integrated circuit enables the successful implementation of the swap test algorithm, offering the capability to prepare not only separable input states but also entangled ones.

\textcolor{black}{
Neglecting the third qubit, in the circuit we have a single photon propagating through a system of four waveguides that allows to represent any generic two-qubit state. Indeed, we mentioned that by associating each path mode (i.e., each state of the qudit) with a computational basis state of the two qubits we obtain a natural isomorphism between the four-dimensional single-photon path-encoded Hilbert space $\mathds{C}^4$ and the two-qubit Hilbert space $\mathds{C}^2 \otimes \mathds{C}^2$. While this mapping allows for the simulation of arbitrary two-qubit states, including those that are entangled from a purely \textit{mathematical} point of view, it is important to stress that no \textit{physical} entanglement is present in this system. Genuine entanglement requires a composite quantum system composed of at least two independently addressable subsystems, each with its own Hilbert space, such that the total state cannot be written as a product state of those subsystems. Furthermore, since there is no canonical or unique way to assign the individual waveguides to the computational basis vectors, the classification of a given state as entangled or separable can depend on the particular choice of encoding. This ambiguity does not arise when the two independent parties of the system correspond to different particles or different degrees of freedom of a single particle (e.g., momentum and polarization of a photon). In our case, however, the system consists of a single photon distributed across four spatially separated waveguides. Since all transformations and measurements are performed on this single photon, the physical degrees of freedom do not support the tensor-product structure required for entanglement in the strict quantum mechanical sense. In other words, it is possible to generate superpositions that mathematically correspond to entangled states, like
\begin{equation}
    \ket{\psi} = \frac{1}{\sqrt{2}}(\ket{1} + \ket{2}) \leftrightarrow \frac{1}{\sqrt{2}}(\ket{01}+\ket{10})
\end{equation}
but no physical entanglement is present. In particular, it is not possible to perform local operations or measurements on independent subsystems as the state is encoded in a single photon. Nevertheless, the photonic state can simulate entangled-state behavior and used for proof-of-concept experiments and practical applications.}

\subsection{The role of $\text{PS}_3$}
From an experimental point of view, the probabilities $\mathds{P}(0)$ and $\mathds{P}(1)$ are estimated by sampling the photon counts at the even- and odd-indexed outputs of the circuit
\begin{equation}
    \mathds{P}(x) = \frac{N_{1-x}}{N_0 + N_1} \quad (x = 0,1)\,,
\end{equation}
where $N_0$ and $N_1$ are the total counts coming from waveguides corresponding to the auxiliary qubit in state $\ket{0}$ and $\ket{1}$ respectively. The number of required detectors can be reduced from eight to four by appropriately tuning the phases of the phase shifters in PS$_3$ and sampling only the even outputs, at the cost of doubling the acquisition time. Indeed, as explained in detail in \cite{baldazzi2024linear}, the phase shifters present in the swap test stage allows not only to counterbalance spurious phases but also to exchange the four outputs corresponding to the state $\ket{0}$ of the ancilla with the four outputs corresponding to the state $\ket{1}$. Therefore, in two runs, it is possible to determine both $\mathds{P}(0)$ and $\mathds{P}(1)$. 

\section{Non-idealities of the circuit}
\subsection{Effect of the non-idealities on the separability threhsold}
\label{app:non_idealities}
Let us assume that the generic two-qubit state $\ket{\Phi} = \alpha \ket{00} + \gamma \ket{10} + \delta \ket{01} + \beta \ket{11}$, with $|\alpha|^2 + |\beta|^2 + |\gamma|^2 + |\delta|^2 = 1$, is set through the preparation stage of the circuit. For the witnessing task, it is important to understand how the non-idealities of the swap test section of the circuit affect the final value of the probability $\mathds{P}(1)$. In particular, we focus on the following non-idealities:
\begin{itemize}
    \item \textbf{Non-ideal MMIs:} due to fabrication errors, the multimode interferometers are not exactly implementing the matrix of an ideal $50:50$ beam splitter. The corresponding matrix representation is modeled as
\begin{equation}
    U_{\text{MMI}}(t,r) = \begin{pmatrix}
    \label{eq:MMI_not_ideal}
t& \text{i}r \\
\text{i}r & t 
\end{pmatrix}\,,
\end{equation}
where $t \neq r \neq \frac{1}{\sqrt{2}}$ and, in general $t^2+r^2 \leq 1$. A spectral characterization of the MMIs was performed using a supercontinuum laser source in combination with an optical spectrum analyzer. The results were obtained by averaging repeated measurements conducted across four different chips. At the experimental operating wavelength of $750$ nm, the normalized transmission and reflection coefficients were measured to be $t^2 = 0.48 (2)$ and $r^2 = 0.52 (2)$, respectively. These coefficients are assumed to be the same for every MMI in the circuit;
\item \textbf{Phase errors:} calibration inaccuracies\footnote{The calibration of the phase shifters in the PIC is itself imperfect due to uncertainties in the measured data, as well as slow drifts caused by temperature fluctuations and power supply instability. These calibration errors can vary between different calibration sets and directly translate into phase errors.}, power supply errors and temperature fluctuations lead to deviations from the exact values of the phases to which every PS is nominally set. These phase errors are incorporated by modifying the ideal unitary for each phase shifter. We define the modified unitary for a phase shifter acting on $n$ waveguides as
\begin{equation}
    U_{\text{PS}n}(\boldsymbol{\theta}, \boldsymbol{\delta}) = \begin{pmatrix}
        {\rm e}^{\text{i}(\theta(1) + \delta(1))} & \dots & 0 \\
        \vdots & \ddots & \vdots \\
        0 & \dots & {\rm e}^{\text{i}(\theta(n) + \delta(n))}
    \end{pmatrix}\,,
\end{equation}
where $\boldsymbol{\theta}$ are the ideal phases and $\boldsymbol{\delta}$ are the associated deviations. Note that the terms $\boldsymbol{\delta}$ do not consider the thermal crosstalk effects (which could introduce additional correlated phase error terms), as it is challenging to quantify their effect on the different PSs; 
\item \textbf{Non-ideal CRs:} The waveguide crossings implementing the CSWAP gate are not described by a perfectly unitary transfer function. The crossing region in the swap test can be modelled as
\begin{equation}
    U_{\text{CWAP}}(T) = \sqrt{T} \begin{pmatrix}
1 & 0 & 0 & 0 & 0 & 0 & 0 & 0 \\
0 & 1 & 0 & 0 & 0 & 0 & 0 & 0 \\
0 & 0 & 1 & 0 & 0 & 0 & 0 & 0 \\
0 & 0 & 0 & 0 & 0 & 1 & 0 & 0 \\
0 & 0 & 0 & 0 & 1 & 0 & 0 & 0 \\
0 & 0 & 0 & 1 & 0 & 0 & 0 & 0 \\
0 & 0 & 0 & 0 & 0 & 0 & 1 & 0 \\
0 & 0 & 0 & 0 & 0 & 0 & 0 & 1 
\end{pmatrix}\,,
\end{equation}
where $T<1$ is the power transmission coefficient of a single crossing component. For our purpose, this non-ideality can be safely neglected under the assumption that all optical paths experience equal losses\footnote{The assumption is justified by the chip design.}. Indeed, if the photon loss is uniform across channels, it does not affect the value of $\mathds{P}(1)$, although it increases the acquisition time. 
\end{itemize}
Let us denote by $\boldsymbol{\theta}_s$ the nominal phases set in the phase shifters of the layer PS$_3$, and by $\boldsymbol{\delta}$ the corresponding errors. Then the output state of the circuit is expressed as
\begin{align}
    \ket{\Psi_{\text{out}}} &= (\mathds{1} \otimes \mathds{1} \otimes U_{\text{MMI}}) \cdot U_{\text{PS}_3}(\boldsymbol{\theta}_s) \cdot U_{\text{CSWAP}} \cdot (\mathds{1} \otimes \mathds{1} \otimes U_{\text{MMI}}) \ket{\Psi} =  \nonumber \\ &= \begin{pmatrix}
         \alpha \left[t^2 {\rm e}^{\text{i} (\theta_s(1) + \delta(1))}-r^2 {\rm e}^{\text{i} (\theta_s(2) + \delta(2))}\right] \\ 
         itr \ \alpha \left[{\rm e}^{\text{i} (\theta_s(1) + \delta(1))}+{\rm e}^{\text{i} (\theta_s(2) + \delta(2))}\right] \\
          \delta t^2 {\rm e}^{\text{i} (\theta_s(3) + \delta(3))} - \gamma r^2 {\rm e}^{\text{i} (\theta_s(4) + \delta(4))}\\
         itr\left[\delta {\rm e}^{\text{i} (\theta_s(3) + \delta(3))} + \gamma e^{\text{i} (\theta_s(4) + \delta(4))}\right] \\
         \gamma t^2{\rm e}^{\text{i} (\theta_s(5) + \delta(5))} - \delta r^2 {\rm e}^{\text{i} (\theta_s(6) + \delta(6))} \\
         itr \left[\gamma {\rm e}^{\text{i} (\theta_s(5) + \delta(5))} + \delta {\rm e}^{\text{i} (\theta_s(6) + \delta(6))}\right] \\
         \beta \left[t^2e^{\text{i} (\theta_s(7) + \delta(7))}-r^2e^{\text{i} (\theta_s(8) + \delta(8))}\right] \\ 
         itr \ \beta \left[{\rm e}^{\text{i} (\theta_s(7) + \delta(7))}+{\rm e}^{\text{i} (\theta_s(8) + \delta(8))}\right] \\
    \end{pmatrix}\,.
\end{align}
and the probability of measuring the ancilla in state $\ket{1}$ is
\begin{align}
    \mathds{P}(1) &= |\alpha|^2 |t^2 {\rm e}^{\text{i}\delta(1)}-r^2{\rm e}^{\text{i}\delta(2)}|^2 + |\beta|^2 |t^2 {\rm e}^{\text{i}\delta(7)}-r^2{\rm e}^{\text{i}\delta(8)}|^2 + \nonumber \\
    & \quad + |\delta t^2 {\rm e}^{\text{i}\delta(3)} - \gamma r^2 {\rm e}^{\text{i}\delta(4)} |^2 + |\gamma t^2 {\rm e}^{\text{i}\delta(5)} - \delta r^2 {\rm e}^{\text{i}\delta(6)}|^2\,.
    \end{align}
    Without loss of generality we can assume that $\gamma \in \mathds{R}^+$ and write $\delta = |\delta|{\rm e}^{\text{i} \phi_{\delta}}$. Then, with some algebra, we obtain
\begin{align}
    \mathds{P}(1)     &= |\alpha|^2 (t^4 +r^4 - 2 r^2 t^2 \cos{\delta_{21}}) +  |\beta|^2 (t^4 +r^4 - 2 r^2 t^2 \cos{\delta_{87}}) + \nonumber \\
    & \quad + (t^4 + r^4)(|\gamma|^2 + |\delta|^2) - 2 |\gamma||\delta| r^2 t^2 \left[\cos{(\phi_{\delta}-\delta_{43})} + \cos{(\phi_{\delta}+\delta_{65})} \right]\,,
\end{align}
where we have introduced the variables $\delta_{ij} \equiv \delta(i) - \delta(j)$ and, in order to keep into account errors and fluctuations, we assume that they are random variables normally distributed with mean $0$ and a standard deviation of $\sigma =0.1$ rad.\\
We are now concerned with the computation of the maximum of $\mathds{P}(1)$ over the set of separable states, associated to the parameter set $A$ defined as
\begin{equation}A=\{(\alpha,\beta,\gamma,\delta)\in \mathds{C}^4:|\alpha\beta-\gamma\delta|^2=0,|\alpha|^2+\be^2 +\ga^2 +\de^2=1  \}\end{equation}
and over all possible realizations of the random variables $\delta_{ij}$ in an interval $[-\sigma,\sigma]$.
Going back to $\mathds{P}(1)$ and considering the maximum over all possible values of $\phi_{\delta}\in [0,2\pi] $, and $\delta_{21}, \delta_{87}, \delta_{65}, \delta_{43}\in [-\sigma,\sigma] $, we can write
\begin{align}
   \max_A \mathds{P}(1) &\leq \max_{A}\Big[|\alpha|^2 (t^4 +r^4 - 2 r^2 t^2 \cos{\sigma}) +  |\beta|^2 (t^4 +r^4 - 2 r^2 t^2 \cos{\sigma}) + \nonumber\\
    & \quad + (t^4 + r^4)(|\gamma|^2 + |\delta|^2) + 4 |\gamma||\delta| r^2 t^2\Big]\nonumber\\
    &=\max_{A}\Big(c|\alpha|^2  + c |\beta|^2  + 
    (t^4 + r^4)(|\gamma|^2 + |\delta|^2) + 4 |\gamma||\delta| r^2 t^2\Big)\,,
\end{align}
where $c = (t^4 +r^4 - 2 r^2 t^2 \cos{\sigma})$. In this case one finds that the maximum point is attained for $|\alpha|=|\beta|=|\gamma|=|\delta| =1/2$ and it is equal to
\begin{equation}\frac{c}{2}+\frac{t^4+r^4+2t^2r^2}{2}=\frac{c}{2}+\frac{(t^2+r^2)^2}{2}=\frac{c}{2}+\frac{1}{2}\,\end{equation}
where in the last equality $t^2+r^2=1$ has been assumed.
\subsection{$2\sigma$ confidence interval on theoretical expectations}
\label{app:non_idealities_2sigma}
When comparing the measured values of the probability $\mathds{P}(1)$ with theoretical expectations, the effect of non-idealities must be considered in both the preparation stage and the swap stage of the circuit. Specifically, for a given set of nominal phases used in the state preparation, we simulate the imperfections of the integrated optical elements by sampling the transmission ($t$) and reflection ($r$) coefficients of the MMIs from Gaussian distributions centered on their experimentally measured means, with corresponding measured variances. Similarly, additional phase errors $\delta$, introduced by imperfections in the phase shifters, are modeled by sampling from a Gaussian distribution with a zero mean and a standard deviation of $0.1$ rad, based on experimental characterization. Using this model, we numerically compute the expected output value of $\mathds{P}(1)$ by simulating the circuit $10 \ 000$ times, each with randomly sampled imperfections\footnote{Only realizations satisfying $t^2+r^2 \leq 1$ are considered.}. This yields the $2\sigma$ confidence interval reported in Tab. \ref{tab:results_bell_2sigma} for the expected values of $\mathds{P}(1)$.
\begin{table}
\renewcommand{\arraystretch}{1.5}
\centering
\caption{Theoretical predictions, $2 \sigma$ confidence intervals (CI), and experimental results for the values of $\mathds{P}(1)$ for the two-qubit Bell states defined in Eq. \eqref{eq:bell_states}. The CIs are computed taking into account the non-idealities of the PIC. }
\label{tab:results_bell_2sigma}
\begin{tabular}{ c c c c }
\toprule
\ &  $\ket{\Phi^{\pm}}$ & $\ket{\Psi^+}$ &  $\ket{\Psi^-}$ \\
\midrule
Theoretical $\mathds{P}(1)$ & 0 & 0 & 1   \\
$2 \sigma$ CI & [0.00\,,\,0.01] & [0.00\,,\,0.04] & [0.84\,,\,1.00]  \\
Experimental $\mathds{P}(1)$ & 0.013(2) & 0.016(4) & 0.98(2)   \\
\bottomrule
\end{tabular}
\end{table}
 \section{Experimental setup}
\label{app:setup}
\begin{figure}
    \centering
    \includegraphics[width=\linewidth]{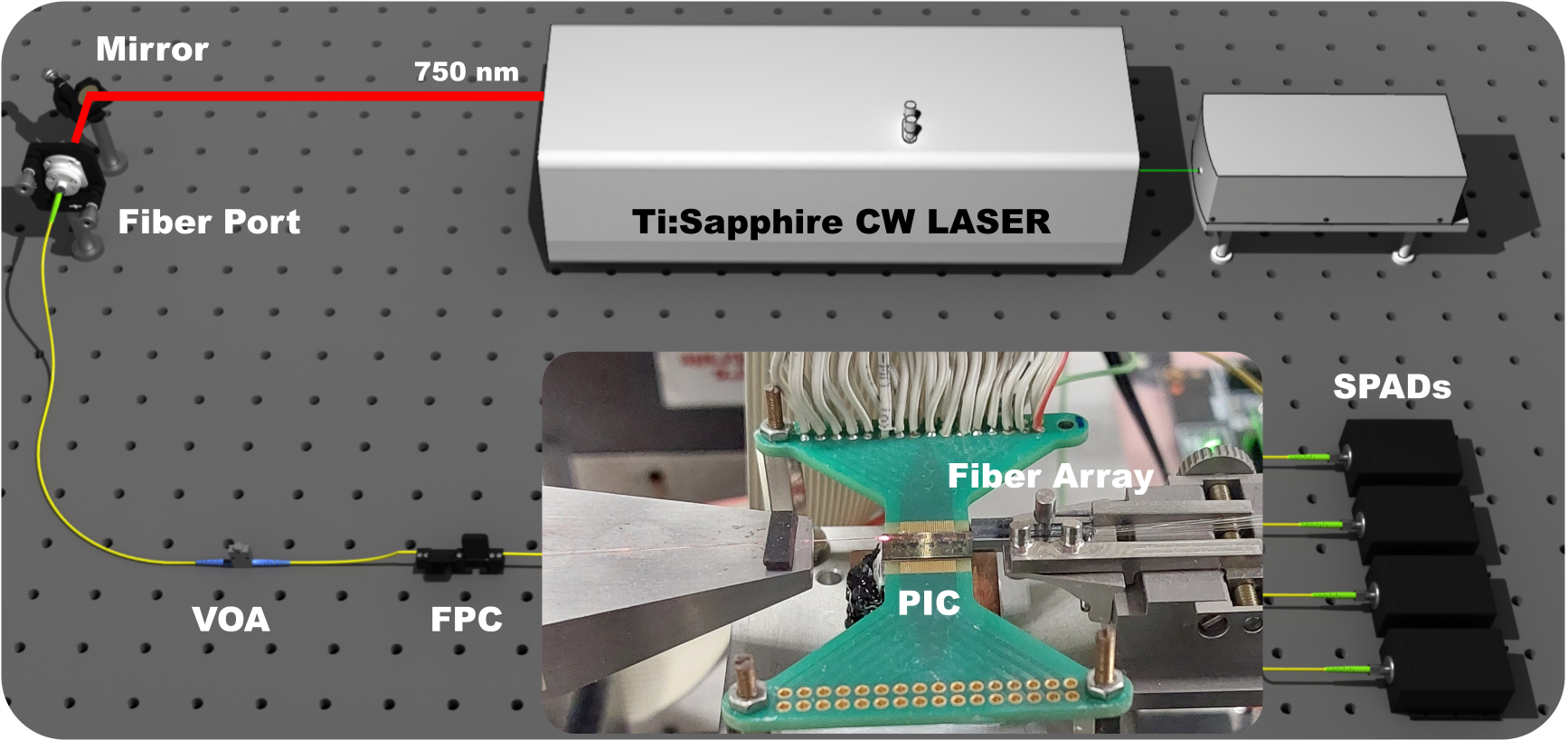}
    \caption{Scheme of the experimental setup (not to scale). A Ti.sapphire continuous-wave laser, tuned to $750$ nm, is used as the light source and coupled into the chip via a lensed tapered optical fiber. The laser beam is attenuated using a variable optical attenuator (VOA), and its polarization is set to be TE at the chip input using a fiber polarization controller (FPC). The output signals are collected via a standard fiber array and directed to four silicon-based single photon avalanche diodes (SPADs) for detection. The photonic chip is mounted on a copper pillar bonded to a Peltier cell, which is regulated by a PID controller for precise temperature control and stabilization. A photograph of the chip is shown in the figure. The PIC is wire-bonded to a printed circuit board (PCB), which is connected to a Qontrol current driver system (Q8iV driver modules) for tunable current delivery to the on-chip phase shifters.}
    \label{fig:exp_setup}
\end{figure}
A schematic of the experimental setup is shown in Fig. \ref{fig:exp_setup}. A Ti:Sapphire laser tuned to a wavelength of $750$ nm is used as the light source. The laser intensity is regulated using a variable optical attenuator (VOA), and its polarization state is adjusted with a fiber polarization controller (FPC) to ensure transverse-electric (TE) polarization at the chip input. Light is injected into the PIC through a lensed tapered fiber, while a standard fiber array collects the optical signal at the output. Subsequently, four identical silicon single-photon avalanche diodes (SPADs, Excelitas) detect the collected photons. The detection events are recorded using a time-tagging module (Swabian Instruments) and transferred to a PC for data acquisition and processing.
\newpage
\printbibliography
\end{document}